\documentclass[english]{article}

\usepackage{algorithmic}
\usepackage{algorithm}
\usepackage{amsfonts, amsmath, amssymb, amsthm}
\usepackage{array}
\usepackage{arydshln}
\usepackage{bm}
\usepackage{color}
\usepackage{comment}
\usepackage{dsfont}
\usepackage{enumitem}
\usepackage{float}
\usepackage[T1]{fontenc}
\usepackage{geometry}
\usepackage{graphicx,multirow}
\usepackage{grffile}
\usepackage{hyperref}\hypersetup{colorlinks,linkcolor=red,anchorcolor=blue,citecolor=blue}
\usepackage[latin9]{inputenc}
\usepackage{letltxmacro}
\usepackage{mathrsfs}
\usepackage{mathtools,natbib}
\usepackage{multirow,xspace,booktabs}
\usepackage{nicefrac}
\usepackage{tablefootnote}
\usepackage{tikz}
\usepackage{verbatim}
\usepackage{subcaption}

\newcommand{\bw}{  {w}}
\newcommand{\bx}{  {x}}
\newcommand{\by}{  {y}}
\newcommand{\bz}{  {z}}

\newcommand{\bB}{  {B}}

\newcommand{\bG}{  {G}}

\newcommand{\bI}{  {I}}

\newcommand{\bM}{  {M}}

\newcommand{\bX}{  {X}}
\newcommand{\bY}{  {Y}}
\newcommand{\bZ}{  {Z}}

\newcommand{\bepsilon}{  {\epsilon}}

\newcommand{\bxi}{  {\xi}}

\newcommand{\defeq}{\coloneqq}

\newcommand{\cA}{\mathcal{A}}

\newcommand{\cF}{\mathcal{F}}

\newcommand{\cK}{\mathcal{K}}
\newcommand{\cL}{\mathcal{L}}
\newcommand{\cM}{\mathcal{M}}
\newcommand{\cN}{\mathcal{N}}

\newcommand{\CC}{\mathbb{C}}

\newcommand{\EE}{\mathbb{E}}

\newcommand{\RR}{\mathbb{R}}

\newcommand{\rme}{\mathrm{e}}
\newcommand{\rmd}{\mathrm{d}}

\newcommand{\argmin}{\mathop{\mathrm{argmin}}}

\newcommand{\TV}{\mathsf{TV}}

\allowdisplaybreaks
\makeatletter 
\let\hat\widehat
\let\epsilon\varepsilon

\theoremstyle{plain}\newtheorem{lemma}{\textbf{Lemma}} 

\newtheorem{theorem}{\textbf{Theorem}}
 
\newtheorem{assumption}{\textbf{Assumption}}

\theoremstyle{definition}
 
\theoremstyle{remark}\newtheorem{remark}{\textbf{Remark}}

\definecolor{yc}{RGB}{255,0,0}
\definecolor{xingyu}{RGB}{0,150,0}

\newcommand{\myalg}{\textsf{\upshape DPnP}}
\newcommand{\DiffSampler}{\textsf{\upshape DDS}}
\newcommand{\GDSampler}{\textsf{\upshape PCS}}
\newcommand{\DDIM}{\textsf{\upshape DDIM}}
\newcommand{\DDPM}{\textsf{\upshape DDPM}}
\newcommand{\cont}{\mathsf{cont}}
\newcommand{\rev}{\mathsf{rev}}
\newcommand{\eqd}{\overset{(\rmd)}{=}}

\newcommand{\acc}{\mathsf{acc}}
\DeclareMathOperator{\Prox}{Prox}

\newcommand{\circo}{\mathbin{\mathchoice
{\xcirc\scriptstyle}
{\xcirc\scriptstyle}
{\xcirc\scriptscriptstyle}
{\xcirc\scriptscriptstyle}
}}
\newcommand{\xcirc}[1]{\vcenter{\hbox{$#1\circ$}}}

\newcolumntype{M}[1]{>{\centering\arraybackslash}m{#1}}

\begin{document}
\title{Provably Robust Score-Based Diffusion Posterior Sampling \\ for Plug-and-Play  Image Reconstruction}

 \author
 {
 	 Xingyu Xu\thanks{Department of Electrical and Computer Engineering, Carnegie Mellon University, Pittsburgh, PA 15213; Emails:
 	 	\texttt{\{xingyuxu,yuejiec\}@andrew.cmu.edu}.} \\
      Carnegie Mellon University
	 	\and
 	 Yuejie Chi\footnotemark[1]  \\
     Carnegie Mellon University
 }

\date{March 2024; Revised May 2024}

\setcounter{tocdepth}{2}
\maketitle

\begin{abstract}
In a great number of tasks in science and engineering, the goal is to infer an unknown image from a small number of noisy measurements collected from a known forward model describing certain sensing or imaging modality. Due to resource constraints, this image reconstruction task is often extremely ill-posed, which necessitates the adoption of expressive prior information to regularize the solution space. Score-based diffusion models, thanks to its impressive empirical success, have emerged as an appealing candidate of an expressive prior in image reconstruction. In order to accommodate diverse tasks at once, it is of great interest to develop efficient, consistent and robust algorithms that incorporate {\em unconditional} score functions of an image prior distribution in conjunction with flexible choices of forward models.

This work develops an algorithmic framework for employing score-based diffusion models as an expressive data prior in nonlinear inverse problems with general forward models. Motivated by the plug-and-play framework in the imaging community, we introduce a diffusion plug-and-play method (\myalg) that alternatively calls two samplers, a  proximal consistency sampler based solely on the likelihood function of the forward model, and a  denoising diffusion sampler based solely on the score functions of the image prior. The key insight is that denoising under white Gaussian noise can be solved {\em rigorously} via both stochastic (i.e., DDPM-type) and deterministic (i.e., DDIM-type) samplers using the same set of score functions trained for generation. We establish both asymptotic and non-asymptotic performance guarantees of \myalg, and provide numerical experiments   to illustrate its promise in solving both linear and nonlinear image reconstruction tasks. To the best of our knowledge, \myalg~is the first provably-robust posterior sampling method for nonlinear inverse problems using unconditional diffusion priors. Code is available at \url{https://github.com/x1xu/diffusion-plug-and-play}.
\end{abstract}

\medskip
\noindent\textbf{Keywords:} score-based generative models, nonlinear inverse problems, posterior sampling, plug-and-play

\tableofcontents{}

\section{Introduction}
\label{sec:intro}

In a great number of sensing and imaging applications, the paramount goal is to infer an unknown image $x^{\star}\in\mathbb{R}^d$ from a collection of measurements $y\in\mathbb{R}^m$ that are possibly noisy, incomplete, and even nonlinear, given as
\begin{equation*}
    y = \mathcal{A}(x^\star) + \xi, 
\end{equation*}
where $\mathcal{A}:\RR^d\to\RR^m$ is the measurement operator encapsulating the {\em forward model} of the sensing modality, and $\xi \in\mathbb{R}^m$ denotes the measurement noise. Examples include restoration tasks such as inpainting, super-resolution, denoising, as well as imaging tasks such as magnetic resonance imaging \citep{lustig2007sparse}, optical imaging \citep{shechtman2015phase}, microscopy imaging \citep{huang2017super}, radar and sonar imaging \citep{potter2010sparsity}, and many more.

Due to sensing and resource constraints, the problem of image reconstruction is often ill-posed, where the desired resolution of the unknown image overwhelms the set of available observations. Consequently, this necessitates the need of incorporating prior information regarding the unknown image to assist the reconstruction process.
 Over the years, numerous types of prior information have been considered and adopted,  from hand-crafted priors such as subspace or sparsity constraints \citep{donoho2006compressed,candes2012exact}, to data-driven ones prescribed in the form of neural networks \citep{ulyanov2018deep,bora2017compressed}. These priors can be regarded as some sort of generative models for the unknown image, which postulate the high-dimensional image admits certain parsimonious representation in a low-dimensional data manifold. It is desirable that the generative models are sufficiently expressive to capture the diversity and structure of the image class of interest, yet nonetheless, still lead to image reconstruction problems that are computationally tractable.

\paragraph{Score-based diffusion models as an image prior.} Recent years have seen tremendous progress on generative artificial intelligence (AI), where it is possible to generate new data samples --- such as images, audio, text --- at unprecedented resolution and scale from a target distribution given training data. Diffusion models, originally proposed by \citet{sohl2015deep}, are among one of the most successful frameworks, underneath popular content generators such as DALL$\cdot$E \citep{ramesh2022hierarchical}, Stable Diffusion \citep{rombach2022high}, Imagen \citep{saharia2022photorealistic}, and many others. Roughly speaking, score-based diffusion models convert noise into samples that resemble those from a target data distribution, by forming the reverse Markov diffusion process only using the score functions of the data contaminated at various noise levels \citep{song2019generative,ho2020denoising,song2020score,song2020denoising}. In particular, \citet{song2020denoising} developed a unified framework to interpret  score-based diffusion models as reversing certain Stochastic Differential Equations (SDE) using either SDE or probability flow Ordinary Differential Equations (ODE), leading to stochastic (i.e., DDPM-type) and deterministic (i.e., DDIM-type) samplers, respectively.  While the DDIM-type sampler is more amenable to acceleration, the DDPM-type sampler tends to generate images of higher quality and diversity when running for a large number of steps \citep{song2020denoising}.
% In contrast with the \DDIM{} sampler in unconditional score-based diffusion sampling, \DDPM{} has a stochastic update equation, which injects Gaussian noise at each step. This extra randomness makes it harder to accelerate \DDPM{}-type samplers, but if the image quality is of primary concern, it was observed that \DDPM{} generates images of higher quality than that of \DDIM{} when run for a large number of steps \citep{song2020denoising}. 

Thanks to the expressive power of score-based diffusion models in generating complex and fine-grained images, they have emerged as a plausible candidate of an expressive  prior in image reconstruction \citep{song2021solving,chung2023diffusion,feng2023score} via the lens of {\em Bayesian posterior sampling}. To accommodate diverse applications with various image characteristics and imaging modalities, it is desirable to develop {\em plug-and-play} methods that do not require training from scratch or end-to-end training for every new imaging task. Nonetheless, despite a flurry of recent efforts, existing algorithms either are computationally expensive \citep{wu2023practical,cardoso2023monte}, inconsistent \citep{chung2023diffusion,kawar2022denoising}, or confined to linear inverse problems \citep{cardoso2023monte,dou2024diffusion}. Therefore, a natural question arises:
\begin{center}
{\em Can we develop a practical, consistent and robust algorithm that incorporates score-based diffusion models as an image prior with general (possibly nonlinear) forward models? }
\end{center}

\subsection{Our contribution}

This paper provides an affirmative answer to this question, by developing an algorithmic framework to sample from the posterior distribution of images, where score-based diffusion models are employed as an expressive image prior in nonlinear inverse problems with general forward models. 
Specifically, our contributions are as follows.  
\begin{itemize}

\item {\em Diffusion plug-and-play for posterior sampling.} Motivated by the plug-and-play \citep{venkatakrishnan2013plug} framework in the imaging community, we introduce a diffusion plug-and-play method (\myalg) that alternatively calls two samplers, a {\em proximal consistency sampler} that aims to generate samples that are more consistent with the measurements, and a {\em denoising diffusion sampler} that focuses on sampling from the posterior distribution of an easier problem --- image denoising under white Gaussian noise --- to enforce the prior constraint. Our method is {\em modular}, in the sense that  the {\em proximal consistency sampler} is solely based on the likelihood function of the forward model, and the denoising diffusion sampler is based solely on the score functions of the image prior. 

\item {\em Posterior sampling for image denoising.} While the proximal consistency sampler can be borrowed somewhat straightforwardly from existing literature such as the Metropolis-adjusted Langevin algorithm \citep{roberts1998optimal}, the denoising diffusion sampler, on the other hand, has not been addressed in the literature to the best of our knowledge. Our key insight is that this can be solved via both stochastic (i.e., DDPM-type) or deterministic (i.e., DDIM-type) samplers by carefully choosing the forward SDEs and discretizing the resulting reversal SDE or ODE using the exponential integrator \citep{zhang2022fast}. Importantly, the denoising diffusion samplers use the same set of unconditional score functions for generation, making it readily implementable without additional training.  

% in terms of the spectral gap of the associated Markov chain

\item {\em Theoretical guarantees.} We establish both asymptotic and non-asymptotic performance guarantees of the proposed \myalg{} method. Asymptotically, we verify the correctness of our method by proving that \myalg{} converges to the conditional distribution of $\bx^\star$ given measurements $\by$, under a suitable choice of annealing schedule and assuming exact unconditional score functions of the image prior.  We next establish a non-asymptotic convergence theory of \myalg{}, where its performance degenerates gracefully with respect to the errors of the samplers, due to, e.g., score estimation errors and limited sampling steps. To the best of our knowledge, this provides the {\em first provably-robust} method for {\em nonlinear} inverse problems using unconditional score-based diffusion priors.%, in terms of the spectral gap of the associated Markov chain.  (?) \yc{check}.
%the general case where the score function estimate, hence the denoising diffusion sampler, can be inaccurate, by establishing a non-asymptotic result of the convergence rate with an explicit dependence on the errors of the denoising diffusion sampler. 
 
\end{itemize}
We further provide numerical experiments to illustrate its promise in solving both linear and nonlinear image reconstruction tasks, such as image super-resolution and phase retrieval. Due to its plug-and-play nature, we expect it to be of broad interest to a wide variety of inverse problems.

\subsection{Related works}

Given its interdisciplinary nature, our work sits at the intersection of generative modeling, computational imaging, optimization and sampling. Here, we discuss some works that are most related to ours.

\paragraph{Algorithmic unrolling and plug-and-play image reconstruction.} Composite optimization algorithms, which aim to minimize the sum of a measurement fidelity term and a regularization term promoting desirable solution structures, have been the backbone of inverse problem solvers. To unleash the power of deep learning, \citet{gregor2010learning} advocates the perspective of algorithmic unrolling, which turns an iterative algorithm into concatenations of linear and nonlinear layers like in a neural network. \citet{venkatakrishnan2013plug} recognized that the proximal mapping step in many composite optimization algorithms can be regarded as a denoiser or denoising operator with respect to the given prior, and proposed to ``plug in'' alternative denoisers, in particular state-of-the-art deep learning denoisers, leading to a class of popular algorithms known as plug-and-play methods \citep{buzzard2018plug}; see \citet{monga2021algorithm} for a review.

\paragraph{Regularization by denoising and score matching.} \citet{vincent2011connection} pointed out a connection between score matching and image denoising, which is a consequence of the Tweedie's formula \citep{efron2011tweedie}. The regularization by denoising (RED) framework \citep{romano2017little} follows the plug-and-play framework to minimize a regularized objective function, where the regularizer is defined based on the plug-in image denoiser; \citet{reehorst2018regularization} later clarified that the RED framework can be interpreted as score matching by denoising using the Tweedie's formula. \citet{kawar2021stochastic} developed a stochastic image denoiser for posterior sampling of image denoising using annealed Langevin dynamics. \citet{fang2024prior} provided a framework to learn exact proximal operators for inverse problems.

\paragraph{Plug-and-play posterior sampling.} Motivated by the need to characterize the uncertainty, tackling  image reconstruction as posterior sampling from a Bayesian perspective is another important approach. Our method is inspired by the plug-and-play framework but takes on a sampling perspective, exploiting the connection between optimization and sampling \citep{wibisono2018sampling}. Along similar lines, \citet{laumont2022bayesian,bouman2023generative} proposed Bayesian counterparts of plug-and-play for posterior sampling, where they leveraged the connection to score matching for sampling from the image prior, but did not consider score-based diffusion models for the image prior, which is a key aspect of ours; see also \citet{sun2023provable}. \citet{coeurdoux2023plug} extended the split Gibbs sampler \citep{vono2019split} in the plug-and-play framework, and advocated the use of score-based diffusion models such as DDPM \citep{ho2020denoising} for image denoising based on heuristic observations. In contrast, we rigorously derive the denoising diffusion samplers from first principles, unraveling critical gaps from na\"ive applications of the generative samplers to denoising, and offer theoretical guarantees on the correctness of our approach.

\paragraph{Score-based diffusion models as image priors.} Several representative methods for solving inverse problems using score-based diffusion priors alternates between taking steps along the diffusion process and projecting onto the measurement constraint, e.g.,  \citet{chung2023diffusion,kawar2022denoising,song2023solving,chung2023fast,graikos2022diffusion,song2022pseudoinverse}. However, these approaches do not possess asymptotic consistency guarantees.  \citet{song2023loss} proposed to use multiple  Monte Carlo samples to reduce bias.
On the other hand, \citet{cardoso2023monte} developed Monte Carlo guided diffusion methods for Bayesian linear inverse problems which tend to be computationally expensive, and \citet{dou2024diffusion} recently introduced a filtering perspective and applied particle filtering. Although asymptotically consistent, these approaches are limited to linear inverse problems. \citet{trippe2022diffusion,wu2023practical} introduced  sequential Monte Carlo (SMC) algorithms for conditional sampling using unconditional diffusion models that are asymptotically exact. \citet{mardani2023variational} developed a variational perspective that connects to the regularization by denoising framework. \citet{gupta2024diffusion} showed that the worst-case complexity of diffusion posterior sampling can take super-polynomial time regardless of the algorithm in use.

\paragraph{Theory of diffusion models and score matching.} A number of recent papers have studied the non-asymptotic convergence rates of popular diffusion samplers, including but not limited to stochastic DDPM-type samplers \citep{chen2022sampling,chen2023improved,li2023towards,benton2024nearly,tang2024contractive}, deterministic DDIM-type samplers \citep{li2023towards,chen2023probability,chen2023restoration}, and accelerated  samplers \citep{li2024accelerating,li2024consistency}. In addition, the statistical efficiency of score matching has also been investigated \citep{koehler2022statistical,pabbaraju2024provable}. %  \citet{hyvarinen2005estimation}  

\subsection{Paper organization and notation}

The rest of the paper is organized is follows. Section~\ref{subsec:score} provides preliminaries on score-based generative models. Section~\ref{sec:algorithm} provides the problem formulation as well as the developed algorithm, followed by its performance analysis in Section~\ref{sec:analysis}. Numerical experiments are provided in Section~\ref{sec:numerical} to elaborate the performance of the proposed algorithm. Last but not least, we conclude in Section~\ref{sec:discussion}.

\paragraph{Notation.} Let $p_{\bx}$ denote the probability distribution of $\bx$, and $p_{\bx}(\cdot | \by)$ denotes the conditional distribution of $\bx$ given $\by$. We use $X \eqd Y$ to denote random variables $X$ and $Y$ are equivalent in distribution. The matrix $I_d$ denotes an identity matrix of dimension $d$. For two probability distributions with density $p(x)$ and $q(x)$, the total variation distance between them is
\[
    \TV(p, q) \defeq \int |p(x) - q(x)| \rmd x.
\]
The $\chi^2$-divergence of $p$ to $q$ is
\[
    \chi^2(p \,\Vert\, q) \defeq \int \frac{(p(x) - q(x))^2}{q(x)} \rmd x.
\]

% \input{related.tex}

%!TeX root = DPnP.tex

\section{Score-based generative models}
\label{subsec:score}

In this section, we set up the preliminary on diffusion-based generative models, which we will be relying upon to develop our algorithm. We shall define two stochastic processes in $\mathbb{R}^d$: 
\begin{itemize}
	\item[1)] a forward process 
		\begin{equation*}
			x_0 \rightarrow x_1 \rightarrow \cdots \rightarrow x_T
		\end{equation*}
		that starts with samples from the target image distribution and diffuses into a noise distribution (e.g.,  standard Gaussians) by gradually injecting noise into the samples; 
	\item[2)] a reverse process 
		\begin{equation*}
			    \bx_{T}^\rev \rightarrow  \bx_{T-1}^\rev \rightarrow \cdots \rightarrow  \bx_{0}^\rev
		\end{equation*}
		that starts from pure noise (e.g., standard Gaussians) and converts it into samples whose distribution is close to the target image distribution. 
\end{itemize}

\subsection{The forward process and score functions}

Consider the forward Markov process in $\mathbb{R}^d$ that starts with a sample from the data distribution $p_X$, and adds noise over the trajectory according to
\begin{subequations}
\label{eq:forward-process}
\begin{align}
	x_0 &\sim p^{\star},\\
	x_t &= \sqrt{1-\beta_t}\, x_{t-1} + \sqrt{\beta_t} \, w_{t}, \qquad 1\leq t\leq T,
\end{align}
\end{subequations}
where $\{w_t\}_{1\leq t\leq T}$'s are independent standard Gaussian vectors, i.e., $w_t\overset{\mathrm{i.i.d.}}{\sim} \mathcal{N}(0, I_d)$, and $\{\beta_t \in (0,1)\}$  describes the noise-injection rates used in each step. Therefore, we can write $x_t$ equivalently as
\begin{equation}
\label{eqn: discrete OU marginal}
\bx_t \defeq \sqrt{\bar\alpha_t}\, \bx_0  + \sqrt{1-\bar\alpha_t}\,\bepsilon_t, \quad \bepsilon_t \sim\mathcal N(0, \bI_d), \quad t=0,1,\cdots,T.
\end{equation}
Here, $(\bar\alpha_t)_{t=0,1,\cdots,T}$ is the \emph{schedule} of diffusion given by
\begin{align}
	\alpha_t\coloneqq 1 - \beta_t, 
	\qquad \bar{\alpha}_t \coloneqq \prod_{k = 1}^t \alpha_k ,\qquad 1\leq t\leq T.
\end{align}
Clearly, it verifies that
$1 \ge \bar\alpha_0 > \bar\alpha_1 > \cdots > \bar\alpha_T > 0$.
As long as $\bar\alpha_T$ is vanishing, it is easy to observe that the distribution of $x_T$ approaches $\mathcal{N}(0, \bI_d)$.

\paragraph{Score functions.} As will be seen, in order to sample from $p^{\star}$, it turns out to be sufficient to learn the score functions of $p_{\bx_t}$ at each step of the forward process, defined as
\begin{equation}\label{eq:score_definition}
    s^\star_t(\bx) = \nabla\log p_{\bx_t}(\bx), \qquad t=0,1,\cdots,T.
\end{equation}
An enlightening property \citep{vincent2011connection} of the score function is that it can be interpreted as the minimum mean-squared error (MMSE) estimate of $\bepsilon_t $ given $\bx_t = \bx$, fueled by Tweedie's formula:
\begin{equation}\label{eq:score_interpretation}
    s^\star_t( \bx) = -\frac{1}{\sqrt{1 - \bar\alpha_t}} \underbrace{ \EE_{\bx_0\sim p^\star,\, \bepsilon_t \sim\mathcal N(0, \bI_d)}(\bepsilon_t \,|\, \sqrt{\bar\alpha_t} \bx_0 + \sqrt{1 - \bar\alpha_t} \bepsilon_t = \bx) }_{ =: \bepsilon_t^{\star}( \bx)}.
\end{equation}
Consequently, this makes it possible to estimate the score functions via learning to denoise \citep{hyvarinen2005estimation}, by estimating the denoising function $\bepsilon_t^{\star}(\cdot)$, as typically done in practice \citep{ho2020denoising}.

\paragraph{Continuous-time perspective.} To facilitate understanding, it will be convenient to formulate the diffusion process in continuous time. To distinguish from the discrete-time setting, we use capitalized letters like $\bX$ to denote the continuous-time diffusion process, and $\tau$ to denote continuous time parameter. The continuous-time forward diffusion follows the Ornstein-Uhlenbeck (OU) process\footnote{In the literature, other processes such as Variance-Exploding SDE (VE-SDE) are also used. Our theory also applies to these processes with straightforward modifications.}, defined by the Stochastic Differential Equation (SDE) \citep{song2020score}:
\begin{equation}\label{eq:banana}
    \rmd \bX_\tau = -\bX_\tau \rmd \tau + \sqrt{2}\,\rmd\bB_\tau, \quad \tau\ge 0, \quad \bX_0 \sim p^\star,
\end{equation}
where $(\bB_\tau)_{\tau \ge 0}$ is the standard $d$-dimensional Brownian motion. It can be shown that \citep{doob1942brownian, evans2012introduction} the marginal distribution of $\bX_\tau$ for $\tau\ge0$ is
\begin{equation}
\label{eqn: OU marginal}
    \bX_\tau \eqd \rme^{-\tau}\bX_0 + \sqrt{1-\rme^{-2\tau}} \bepsilon, \quad \bX_0 \sim p^\star, ~\bepsilon\sim\mathcal N(0, \bI_d).
\end{equation}
Comparing \eqref{eqn: discrete OU marginal} and \eqref{eqn: OU marginal}, it can be checked that the discrete-time diffusion process can be embedded into the continuous-time one via the time change $t\mapsto \frac12 \log\frac{1}{\bar\alpha_t}$, in the sense that $\bx^\star_t \overset{(\rmd)}{=} \bX_{\frac12 \log\frac{1}{\bar\alpha_t}}$.

Similar to the discrete-time case, the continuous-time score function is defined by
\begin{align}
    s^\cont(\tau, \bx) & = \nabla\log p_{\bX_\tau}(\bx) \label{eq:continuous_score}\\
    & = -\frac{1}{\sqrt{1 - \rme^{-2\tau}}} \underbrace{ \EE_{\bX_0 \sim p^\star,\, \bepsilon\sim\mathcal N(0, \bI_d)}\big( \bepsilon \,|\, \rme^{-\tau} \bX_0 + \sqrt{1 - \rme^{-2\tau}} \bepsilon = \bx \big) }_{=: \bepsilon^\cont(\tau, \bx) } , \nonumber
\end{align}
where the second line follows again from the Tweedie's formula, where $\bepsilon^\cont(\tau, \bx) $ is the continuous-time counterpart of $\bepsilon_t^{\star}(\cdot)$. 
By the change-of-time argument mentioned above, we have the following correspondence between the discrete-time and the continuous-time score functions:
\begin{equation}
\label{eqn: conversion between discrete and continuous score} 
s^\cont \left( \frac12 \log\frac{1}{\bar\alpha_t},\,\bx \right) = s^\star_t( \bx).
\end{equation}

\subsection{The reverse process and sampling}

To enable sampling, one needs to ``reverse'' the forward diffusion process. Fortunately, it is possible to leverage classical theory \citep{anderson1982reverse,ambrosio2005gradient} to reverse the SDE, and apply discretization to the time-reversal processes to collect samples. We shall describe two popular approaches below, corresponding to stochastic (i.e., DDPM-type) and deterministic (i.e., DDIM-type) samplers respectively following primarily the framework set forth in \citet{song2020score}.

\paragraph{Time-reversed SDEs and probability flow ODEs.} Let us begin with the more general theory of {\em reversing} SDEs, which will be useful in future sections. Consider a SDE given by
\begin{equation}
    \label{eqn: forward SDE}
    \rmd \bM_\tau = f(\bM_\tau) \rmd\tau + \sqrt{\beta} \rmd \bB_\tau, \quad \tau\ge 0, \quad \bM_0 \sim p_{\bM_0},
\end{equation}
where $f: \RR^d \to \RR^d$ is a deterministic function and $\beta>0$ is a constant. For any positive time $\tau_\infty>0$, define the reversed time parameter
\begin{equation}
\label{eqn: reverse time parameter}
\tau^\rev \coloneqq \tau^\rev(\tau) = \tau_\infty - \tau.
\end{equation}
We are now ready to describe the time-reversed processes.
\begin{itemize}
	\item[1)] The {\em time-reversed SDE} of \eqref{eqn: forward SDE} on the time interval $[0, \tau_\infty]$ is defined as
\begin{equation}
    \label{eqn: time reversal SDE}
    \begin{gathered}
        \rmd \bM^\rev_{\tau^\rev} 
        = \left( -f( \bM^\rev_{\tau^\rev}) + \beta \nabla \log p_{\bM_{\tau^\rev}}(\bM^\rev_{\tau^\rev}) \right) \rmd\tau 
        + \sqrt{\beta} \rmd \tilde\bB_\tau, \quad 
         \tau\in[0, \tau_\infty], \quad \bM^\rev_{\tau_\infty} \sim p_{\bM_{\tau_\infty}},
    \end{gathered}
\end{equation}
where $\tilde\bB$ is an independent copy of $\bB$, i.e., another Brownian motion.  It is a classical result in stochastic analysis \citep{anderson1982reverse} that the reversed process $\bM^\rev$ shares the same path distribution as $\bG$, i.e., 
\[
    (\bM^\rev_{\tau})_{\tau\in[0, \tau_\infty]} \eqd (\bM_{\tau})_{\tau\in[0, \tau_\infty]}.
\]
In other words, the joint distribution of $(\bM^\rev_{\tau_1}, \bM^\rev_{\tau_2}, \cdots, \bM^\rev_{\tau_k})$ for any $0\le \tau_1\le \tau_2\le \cdots \le \tau_k \le \tau_\infty$, for any integer $k\ge 1$, coincides with that of $(\bM_{\tau_1}, \bM_{\tau_2}, \cdots, \bM_{\tau_k})$. 

\item[2)]  In place of the reversed SDE in \eqref{eqn: time reversal SDE}, it is possible to consider the following probability flow ODE \citep{ambrosio2005gradient,song2020score}:
\begin{equation}
    \label{eqn: time reversal ode}
    \rmd \bM^\rev_{\tau^\rev} = \left( -f( \bM^\rev_{\tau^\rev} )+ \frac{\beta}{2} \nabla \log p_{\bM_{\tau^\rev}}(\tau^\rev, \bM^\rev_{\tau^\rev}) \right) \rmd \tau, \quad \tau\in[0, \tau_\infty], \quad \bM^\rev_{\tau_\infty} \sim p_{\bM_{\tau_\infty}}.
\end{equation}
The reversed ODE satisfies a slightly weaker guarantee than that of the reversed SDE, which nevertheless suffices for most practical purposes \citep{song2020score}:
\[
    \bM^\rev_\tau \eqd \bM_\tau, \quad \tau\in[0, \tau_{\infty}].
\]
Note that the reversed ODE only guarantees identical marginal distribution for each $\bM^\rev_\tau$, whereas the reversed SDE guarantees identical joint distribution. 
%  which is the continuous-time analogy of \eqref{eqn: DDIM guarantee}. 
\end{itemize}

Specializing the above to the OU process \eqref{eq:banana} with proper discretization then leads to popular samplers used for generation, as follows.

\paragraph{\DDPM{}-type stochastic samplers.}
Specializing the time-reversed SDE \eqref{eqn: time reversal SDE} to the OU process gives
\[
    \rmd \bX^\rev_{\tau^\rev} = \big( \bX^\rev_{\tau^\rev} + 2s^\cont(\tau^\rev, \bX^\rev_{\tau^\rev}) \big) \rmd \tau + \sqrt{2} \rmd \tilde\bB_\tau, \quad \tau\in[0, \tau_\infty], \quad \bX^\rev_{\tau_\infty} \sim p_{\bX_{\tau_\infty}}.
\]
As $\tau_\infty\to\infty$, it can be seen from \eqref{eqn: OU marginal} that $p_{\bX_{\tau_\infty}}$ converges in distribution to $\bepsilon\sim\cN(0, \bI_d)$. Thus the solution of the above SDE can be approximated by initializing $\bX^\rev_{\tau_\infty}\sim \cN(0, \bI_d)$ instead. The \DDPM{} sampler \citep{ho2020denoising} can be viewed as a discretization of this SDE \citep{song2020score}.

\paragraph{\DDIM{}-type deterministic samplers.}
On the other hand, the probability flow ODE \eqref{eqn: time reversal ode} for the OU process reads as
\begin{equation}
    \label{eqn:ode}
    \rmd \bX^\rev_{\tau^\rev} = \big( \bX^\rev_{\tau^\rev} + s^\cont(\tau^\rev, \bX^\rev_{\tau^\rev}) \big) \rmd \tau, \quad \tau\in[0, \tau_\infty], \quad \bX^\rev_{\tau_\infty} \sim p_{\bX_{\tau_\infty}}.
\end{equation}
Again, as $\tau_\infty \to \infty$, one may approximate the initialization with $\bX^\rev_{\tau_\infty} \sim \cN(0, \bI_d)$. It is known that the popular \DDIM{} sampler \citep{song2020score,song2020denoising} is a discretization of this ODE \citep{zhang2022fast}. The ODE-based deterministic samplers allow more aggressive choice of discretization schedules, as well as fast ODE solvers \citep{lu2022dpm}, enabling significantly accelerated sampling process compared to the SDE-based stochastic samplers. 

\begin{remark} 
In correspondence with the discrete-time schedule, one might take $\tau_{\infty}=-\frac12 \log \bar\alpha_T$. We refer interested readers to \citet{song2020score,tang2024score} for more details.
\end{remark}

%!TeX root = DPnP.tex

\section{Posterior sampling via diffusion plug-and-play}
\label{sec:algorithm}

\subsection{Posterior sampling for image reconstruction}
We are interested in solving (possibly nonlinear) inverse problems, where the aim is to infer an unknown image $\bx^\star\in\RR^d$ from its measurements $\by\in\RR^m$,\footnote{For simplicity, we limit our presentation to the real-valued case; our framework generalizes to the complex-valued case in a straightforward manner.} given by
\begin{equation*}
    \by = \cA(\bx^\star) + \bxi, 
\end{equation*}
where $\cA:\RR^d\to\RR^m$ is the measurement operator underneath the forward model, and $\bxi$ denotes measurement noise. It has been well-understood that \emph{prior information} of $\bx^\star$ plays an important role in solving inverse problems that are otherwise ill-posed,  enabling successful reconstruction with much less measurements and higher accuracy. At the same time, it is desirable to understand and quantify the uncertainty in image reconstruction, especially when the available measurements are rather limited.

\paragraph{Posterior sampling.}
In this work, we focus on the Bayesian setting where the prior information of $\bx^\star$ is provided in the form of  some prior distribution $p^\star(\cdot)$, i.e., 
\begin{align}
	x^{\star} \sim p^{\star}(x),
\end{align}
The \emph{posterior distribution} given measurements $\by$ is defined as 
\begin{equation}
\label{eqn: posterior factorization}
p^\star(\bx|\by) \propto p^\star(\bx) \, p(\by|\bx^\star=\bx) = p^\star(\bx)\, \rme^{\cL(\bx; \by)}.
\end{equation}
Here, $\cL(\cdot; \by)$ is the log-likelihood function of the measurements. For example, when the noise $\xi\sim\mathcal N(0, \sigma^2\bI_m)$ is standard Gaussian, it follows that
\[
    \cL(\bx; \by) = -\frac{1}{2\sigma^2} \|\by - \cA(\bx)\|^2 - \frac m 2 \log(2\pi\sigma^2).
\]
Notwithstanding, our framework allows flexible choices of the forward model and the noise distributions. In addition, while this formulation is derived from probabilistic interpretations, it also subsumes the ``reward-guided'' or ``loss-guided'' setting \citep{song2023loss}, where $\cL$ can be viewed as a reward function or a negative loss function, both of which characterize preference over structural properties of $\bx^\star$. In all these settings, it will be useful to bear in mind the intuition that the higher value of $\cL$ corresponds to better consistency with the measurements, higher rewards, etc.

\paragraph{Assumption on the forward model.} Throughout the paper, for simplicity, we make the following mild assumption on $\cL$, which is applicable to many applications of interest.  

\begin{assumption}
\label{assumption:L}
We assume $\cL(\cdot \,;\, y)$ is differentiable almost everywhere, and $\sup_{x \in \RR^d} \cL(x; y) < \infty$.
\end{assumption}

\paragraph{Goal.} Our goal is to sample $\hat{x}$ from the posterior distribution 
$$\hat{x} \sim p^\star(\cdot\,|\,\by)$$ given estimates $\hat{s}_t(\bx)$ (resp. $\hat\bepsilon_t(\bx)$) of the {\em unconditional} score functions $s^\star_t(\bx)$ (resp. the noise function $ \bepsilon_t^{\star}( \bx)$) in \eqref{eq:score_definition}, assuming knowledge of the likelihood function $\cL(\cdot; \by)$.

\subsection{Key ingredient: score-based denoising posterior sampling}
\label{subsec: dds}

We begin with an inspection on one of the most fundamental inverse problems: denoising under white Gaussian noise. We shall demonstrate how to solve this problem via  stochastic (i.e., DDPM-type) and deterministic (i.e., DDIM-type) denoising diffusion samplers using the same set of unconditional score functions trained for generation. As shall be elucidated shortly, the denoising diffusion samplers turn out to be an important building block in our algorithm for general inverse problems. 

\paragraph{Image denoising under white Gaussian noise.} Suppose that we have access to a noisy version of $\bx^\star \sim p^{\star}$ contaminated by white Gaussian noise, given by
\begin{equation} \label{eq:cookie}
    x_{\mathsf{noisy}} = \bx^\star + \bxi, \quad \bxi\sim\mathcal N(0, \eta^2\bI_d),
\end{equation}
where $\eta>0$ is the noise intensity {\em assumed to be known}. Our goal is to sample from $p^\star(\cdot \,|\, x_{\mathsf{noisy}})$ given the score estimates  $\hat{s}_t(\bx)$ (resp. the noise estimates $\hat\bepsilon_t(\bx)$).  
We will develop our score-based denoising posterior sampler, termed \DiffSampler, with two variants, \DiffSampler-\DDPM{} and \DiffSampler-\DDIM{}, which can be viewed as analogues of the well-known \DDPM{} and \DDIM{} samplers in unconditional score-based sampling respectively. 
Before proceeding, it is worth highlighting that the two variants will be derived from different forward diffusion processes, since we observe the resulting variants empirically lead to more competitive performance.

\paragraph{A stochastic \DDPM{}-type sampler via heat flow.} We begin with a stochastic \DDPM{}-type sampler for denoising, termed \DiffSampler-\DDPM{}. We divide our development into the following steps. 
\begin{itemize}
	\item[1)] {\em Step 1: introducing the heat flow.} Let us introduce a {\em heat flow} with initial distribution $p^\star$, defined by the following SDE:
\begin{equation}
    \label{eqn: heat flow}
    \rmd \bY_\tau = \rmd \bB_\tau,\quad \tau\ge 0, \quad \bY_0 \sim p^\star,
\end{equation}
where $(\bB_\tau)_{\tau\ge 0}$ is the standard $d$-dimensional Brownian motion. The  solution of \eqref{eqn: heat flow} is simply
\begin{equation}
\label{eqn: solution of heat flow}
\bY_\tau = Y_0 + \bB_\tau, \quad \tau \ge 0.
\end{equation}
Since $\bB_\tau \sim \cN(0, \tau I_d)$, it readily follows that $\bB_{\eta^2} \eqd \xi$, which together with $Y_0\sim p^\star$ yield the important observation that $x_{\mathsf{noisy}} = \bx^\star + \bxi$ can be viewed as an endpoint of the heat flow, in the sense that
\[
    x_{\mathsf{noisy}} = \bx^\star + \xi   \eqd \bY_{\eta^2}.
\]

\item[2)] {\em Step 2: reversing the heat flow.} Following similar reasonings in Section~\ref{subsec:score}, the next step boils down to reverse the heat flow \eqref{eqn: heat flow}.
The time-reversal of the heat flow SDE \eqref{eqn: heat flow} is (cf. \eqref{eqn: time reversal SDE}) given by
\begin{equation}
\label{eqn: reverse heat flow}
\rmd \bY^\rev_{\eta^2 - \tau} = \nabla \log p_{\bY_{\eta^2 - \tau}} (\bY^\rev_{\eta^2 - \tau}) \rmd \tau + \rmd \tilde{\bB}_\tau,\quad \tau\in[0,\eta^2], \quad \bY^\rev_{\eta^2} \sim p_{\bY_{\eta^2}},
\end{equation}
where $(\tilde\bB_\tau)_{\tau\ge 0}$ is an independent copy of $(\bB_\tau)_{\tau\ge 0}$. As introduced earlier, the virtue of the time-reversed SDE \eqref{eqn: reverse heat flow} is that it produces a process $\bY^\rev_\tau$ with the same \emph{path} distribution as $\bY_\tau$, i.e., 
\[
    (\bY^\rev_\tau)_{\tau\in[0,\eta^2]} \eqd (\bY_\tau)_{\tau\in[0,\eta^2]}.
\]
In particular, the joint distribution of $(\bY^\rev_0, \bY^\rev_{\eta^2})$ is the same as that of $(\bY_0, \bY_{\eta^2}) \eqd (\bx^\star, x_{\mathsf{noisy}})$. This implies that the conditional distribution $p^\star(\cdot \,|\, x_{\mathsf{noisy}})$ is the same as $p_{\bY^\rev_0}(\cdot \,|\, \bY^\rev_{\eta^2}=x_{\mathsf{noisy}})$. Surprisingly, the latter admits a simple interpretation: $p_{\bY^\rev_0}(\cdot \,|\, \bY^\rev_{\eta^2}=x_{\mathsf{noisy}})$ is the distribution of $\bY^\rev_0$ when we initialize \eqref{eqn: reverse heat flow} with $\bY^\rev_{\eta^2} = x_{\mathsf{noisy}}$! Therefore, sampling the posterior $p^\star(\cdot \,|\, x_{\mathsf{noisy}})$ amounts to solving the following simple SDE:
\begin{equation}
\label{eqn: SDE for DDS}
\rmd \bY^\rev_{\eta^2 - \tau} = \nabla \log p_{\bY_{\eta^2 - \tau}} (\bY^\rev_{\eta^2 - \tau}) \rmd\tau + \rmd \tilde{\bB}_\tau,\quad \tau\in[0,\eta^2], \quad \bY^\rev_{\eta^2} = x_{\mathsf{noisy}}.
\end{equation}

\item [3)] {\em Step 3: connecting the score functions.} It is now immediate to arrive at our proposed stochastic sampler \DiffSampler-\DDPM{} by discretization of this SDE \eqref{eqn: SDE for DDS}, which requires knowledge of the score functions $\nabla \log p_{\bY_\tau}(\cdot)$. A key observation is that they can in fact be computed from the score function $s^\cont(\tau, \bx) $ (cf.~\eqref{eq:continuous_score}), due to the following lemma, whose proof is provided in Appendix~\ref{proof:score_hf}. 
\begin{lemma}[Score function of $\bY_\tau$]
\label{lem: score of heat flow}
For $\tau\ge 0$, we have
\[
    \nabla \log p_{\bY_\tau}(\bx) = \frac{1}{\sqrt{1 + \tau}} s^\cont \left( \frac12 \log(1+\tau),\, \frac{\bx}{\sqrt{1 + \tau}} \right).
\]
\end{lemma}
We leave the detailed discretization procedure with an exponential integrator \citep{zhang2022fast} to Appendix~\ref{subsec: discretizing DDS-DDPM}. The resulting sampler, \DiffSampler-\DDPM{}, is summarized in Algorithm~\ref{alg:DDS-DDPM}. 
\end{itemize}

\begin{algorithm}[ht]
\caption{Denoising Diffusion Sampler (stochastic) $\DiffSampler\text{-}\DDPM(\bx_{\mathsf{noisy}}, \hat s, \eta)$} \label{alg:DDS-DDPM} 
\begin{algorithmic} 
\STATE \textbf{{Input}}: noisy data $\bx_{\mathsf{noisy}}\in\RR^d$,  score estimates $\hat s: = \{ \hat s_t(\cdot):\RR^d\to\RR^d, t=1,\ldots, T\}$ or noise estimates $\hat{\bepsilon} =\{ \hat{\bepsilon}_t(\cdot):\RR^d\to\RR^d, t=1,\ldots, T\}$, and noise level $\eta>0$.
\STATE \textbf{{Scheduling}}: Compute the diffusion schedule $(\tau_t)_{0\le t\le T'}$ by
\[
    \tau_t = \bar\alpha_t^{-1} - 1, \quad 0\le t \le T',
\]
where
\[
    T' \defeq \max\left\{ t: 0\le t \le T,\,\bar\alpha_t > \frac{1}{\eta^2 + 1} \right\}.
\]
\STATE \textbf{{Initialization}}: Set $\hat\bx_{T'} = \bx_{\mathsf{noisy}}$.
\STATE \textbf{{Diffusion}}: \textbf{for} $t=T', T'-1, \ldots, 1$ \textbf{do} 
\[
    \hat\bx_{t-1} = \hat\bx_t - 2(\sqrt{\tau_t} - \sqrt{\tau_{t-1}})\, \hat{\bepsilon}_{t} + \sqrt{\tau_t - \tau_{t-1}}\, \bw_t , \quad \bw_t \sim\cN(0, \bI_d).
\]
where
\begin{align*}
    \hat{\bepsilon}_t &\defeq \hat{\bepsilon}_t(\sqrt{\bar\alpha_t} \, \hat\bx_t) = -\frac{1}{\sqrt{1-\bar\alpha_{t}}} \hat s_t\left(\sqrt{\bar\alpha_t} \,\hat\bx_t \right).
\end{align*}
\STATE \textbf{{Output}}: $\hat\bx_0$.
\end{algorithmic} 
\end{algorithm}

\paragraph{A deterministic \DDIM{}-type sampler via OU process. } 
We next develop a deterministic \DDIM{}-type sampler for denoising, termed \DiffSampler{}-\DDIM{}, presented in Algorithm~\ref{alg:DDS-DDIM}. 

\begin{itemize}
	\item[1)] {\em Step 1: introducing a posterior-initialized OU process.} To sample from the posterior distribution $p^\star(\cdot | x_{\mathsf{noisy}})$, we first introduce a random variable $\bw$ which has (unconditional) distribution  
    \begin{equation}
        \label{eqn: p_w}
        p_{\bw}(\bx) \defeq p^\star(\bx^\star = \bx \,|\, \bx^\star + \xi = \bx_{\mathsf{noisy}}),
    \end{equation}
     in the same form of the desired posterior distribution $p^\star(\cdot|\bx_{\mathsf{noisy}})$. Here, since the noisy observation $x_{\mathsf{noisy}}$ is given, we regard it as fixed.\footnote{Technically, this can done by conditioning on $x_{\mathsf{noisy}}$ throughout our discussion of \DiffSampler-\DDIM.}
    We then further introduce $\bz = \bw - x_{\mathsf{noisy}}$, which is a ``centered'' version of $\bw$, whose distribution is 
    \[
        p_{\bz}(\bx) \defeq p_{\bw} (\bx +\bx_{\mathsf{noisy}}) =  p^\star(\bx^\star = \bx + \bx_{\mathsf{noisy}} \,|\, \bx^\star + \xi = \bx_{\mathsf{noisy}}).
    \]
    The OU process with initial distribution $p_z$ is defined by the SDE:
    \begin{equation}
        \label{eqn: OU denoising def}
        \rmd Z_\tau = -Z_\tau \rmd \tau + \rmd B_\tau, \quad \tau \ge 0, \quad Z_0 \sim p_z,
    \end{equation}
    where $B_\tau$ is the standard $d$-dimensional Brownian motion. As in \eqref{eqn: OU marginal}, the marginal distribution of $Z_\tau$ is given by
	\begin{equation}\label{eqn:bla}
    \bZ_\tau \eqd \rme^{-\tau}\bZ_0 + \sqrt{1 - \rme^{-2\tau}} \bepsilon, \quad \bZ_0 \sim p_z, ~\bepsilon\sim\mathcal N(0, \bI_d), \quad \tau \ge 0.
    \end{equation}
	
	\item[2)] {\em Step 2: reversing the OU process.} Following similar reasonings in Section~\ref{subsec:score}, reversing the OU process \eqref{eqn: OU denoising def} will enable us to generate samples $\bz \sim p_{\bz}$. Then we can set $\bw = \bz + \bx_{\mathsf{noisy}}$, which, by definition, has distribution $p_w$ defined in \eqref{eqn: p_w}, and is a sample from the desired posterior distribution $p^\star(\cdot|\bx_{\mathsf{noisy}})$. We are thus led to solve the time-reversed probability flow ODE (cf. \eqref{eqn: time reversal ode}) of \eqref{eqn: OU denoising def}, given by
	\begin{equation}
        \label{eqn: ODE for DDS}
        \rmd \bZ^\rev_{\tau^\rev} = \big( \bZ^\rev_{\tau^\rev} + \nabla \log p_{\bZ_{\tau^\rev}}(\tau^\rev, \bZ^\rev_{\tau^\rev}) \big) \rmd \tau,\quad\tau\in[0, \tau_\infty], \quad \bZ^\rev_{\tau_\infty}\sim\mathcal N(0, \bI_d), 
        \quad \tau^\rev = \tau_\infty-\tau.
    \end{equation}
	
	\item[3)] {\em Step 3: connecting the score functions.} We are now one step away from our proposed deterministic sampler \DiffSampler-\DDIM{}, which is derived by discretization of the ODE \eqref{eqn: ODE for DDS}. We need to know the score functions $\nabla \log p_{\bZ_{\tau}}(\cdot)$, which again can be computed from the score function $s^\cont(\tau, \bx) $ (cf.~\eqref{eq:continuous_score}), as documented by the following lemma, whose proof is provided in Appendix~\ref{proof:score_OU}.
	\begin{lemma}[Score function of $\bZ_\tau$] For $\tau\ge 0$, we have
    \label{lem: denoising score}
    \begin{equation} % s_{\bZ}(\tau, \bx) \defeq
    \label{eqn: denoising score}
    \nabla \log p_{\bZ_\tau}(\bx) = - \frac{\rme^{2\tau} \bx}{\eta^2 + \rme^{2\tau} - 1} + \frac{\rme^{\tau-\tilde\tau}\eta^2}{\eta^2 + \rme^{2\tau} - 1} s^\cont \left( \tilde\tau, \, \rme^{-\tilde\tau}x_{\mathsf{noisy}} + \frac{\rme^{\tau-\tilde\tau}\eta^2 \bx}{\eta^2 + \rme^{2\tau} - 1} \right),
    \end{equation}
    where
    \begin{equation}
        \label{eqn: tau prime def}
        \tilde\tau \coloneqq \tilde\tau(\tau) = \frac12 \log\left( \frac{\eta^2 (\rme^{2\tau} - 1)}{\eta^2 + \rme^{2\tau} - 1} + 1\right).
    \end{equation}
    \end{lemma}
    After plugging this into \eqref{eqn: ODE for DDS} and solving the ODE for $\bZ_\tau^\rev$, we see that $\bZ_0^\rev + \bx_{\mathsf{noisy}}$ is the desired sample from the posterior distribution $p^\star(\cdot|\bx_{\mathsf{noisy}})$, as argued before. 
    Numerically, the ODE \eqref{eqn: ODE for DDS} is solved by discretization with an exponential integrator \citep{zhang2022fast}, resulting in the sampler \DiffSampler-\DDIM{} as summarized in Algorithm~\ref{alg:DDS-DDIM}. Details of the discretization, which are somewhat nuanced, are left to Appendix~\ref{subsec: discretizing DDS-DDIM}.
\end{itemize}

\begin{algorithm}[ht]
\caption{Denoising Diffusion Sampler (deterministic) $\DiffSampler\text{-}\DDIM(\bx_{\mathsf{noisy}}, \hat s, \eta)$}
\label{alg:DDS-DDIM} 
\begin{algorithmic} 
\STATE \textbf{{Input}}: noisy data $\bx_{\mathsf{noisy}}\in\RR^d$, score estimates $\hat s: = \{ \hat s_t(\cdot):\RR^d\to\RR^d, t=1,\ldots, T\}$ or noise estimates $\hat{\bepsilon} =\{ \hat{\bepsilon}_t(\cdot):\RR^d\to\RR^d, t=1,\ldots, T\}$, and noise level $\eta>0$.
\STATE \textbf{{Scheduling}}: Compute the diffusion schedule $(\bar u_t)_{0\le t\le T'}$ by
\[
    \bar u_t = \frac{(\eta^2+1)\bar\alpha_t - 1}{\eta^2 + \bar\alpha_t - 1}, \quad 0\le t \le T',
\]
where
\[
    T' \defeq \max\left\{ t: 0\le t \le T,\,\bar\alpha_t > \frac{1}{\eta^2 + 1} \right\}.
\]
\STATE \textbf{{Initialization}}: Draw $\bz_{T'}\sim\mathcal N(0,\bI_d)$.
\STATE \textbf{{Diffusion}}: \textbf{for} $t=T', T'-1, \ldots, 1$ \textbf{do} 
\[
    \bz_{t-1} = \frac{\sqrt{(\eta^2-1)\bar u_{t-1}+1}}{\sqrt{(\eta^2-1)\bar u_{t}+1}} \bz_{t} + {\sqrt{(\eta^2-1)\bar u_{t-1}+1}}\cdot \big(h(\eta, \bar u_{t-1}) - h(\eta, \bar u_{t})\big) \hat{\bepsilon}_{t}, 
\]
where
\begin{align*}
    h(\eta, u) &\defeq -\arctan\frac{\eta}{\sqrt{u^{-1} - 1}}, \\
    \hat{\bepsilon}_t &\defeq \hat{\bepsilon}_t\left(\sqrt{\bar\alpha_{t}} \bx_{\mathsf{noisy}} + \frac{\eta^2\sqrt{\bar u_t \bar\alpha_{t}} \bz_{t}}{(\eta^2 - 1)\bar u_{t} + 1}\right) = -\frac{1}{\sqrt{1-\bar\alpha_{t}}} \hat s_t\left(\sqrt{\bar\alpha_{t}} \bx_{\mathsf{noisy}} + \frac{\eta^2\sqrt{\bar u_t \bar\alpha_{t}} \bz_{t}}{(\eta^2 - 1)\bar u_{t} + 1}\right).
\end{align*}
\STATE \textbf{{Output}}: $\bx_{\mathsf{noisy}} + \bz_0$.
\end{algorithmic} 
\end{algorithm}

\subsection{Our algorithm: diffusion plug-and-play} 
Now we turn to the general setting where the measurement operator $\cA$ is arbitrary. From the factorization of posterior distribution in \eqref{eqn: posterior factorization}, one intuitively understands that a posterior sampler must obey two constraints simultaneously: (i) the \emph{data prior constraint}, corresponding to the first factor $p^\star(\bx)$, which imposes that the posterior sampler should be less likely to sample at those points where $p^\star(\bx)$ is small; (ii) the \emph{measurement consistency constraint}, corresponding to the second factor $\rme^{\cL(\bx; \by)}$, which imposes that $\cA(\bx)\approx \by$. 

\begin{algorithm}[t]
\caption{Diffusion Plug-and-Play (\myalg)}\label{alg:DPnP} 
\begin{algorithmic} 
\STATE \textbf{{Input}}: Measurements $\by\in\RR^m$, log-likelihood function $\cL(\cdot;\by)$ of the forward model, score estimates $\hat s(\cdot)$, annealing schedule $(\eta_k)_{0\le k\le K}$.
\STATE \textbf{{Initialization}}: Sample $\hat\bx_0\sim\mathcal N(0, \frac{\eta_0}{4}\bI_d)$
\STATE \textbf{{Alternating sampling}}: \textbf{for} $k=0,1,2,\dots,K-1$ \textbf{do} 
\begin{enumerate}[label=(\arabic*), topsep=0pt, itemsep=0em, partopsep=0pt]
    \item {\em Proximal consistency sampler:} Sample $\hat\bx_{k+\frac12}\propto\exp\left(\cL(\cdot \,;\, \by) - \frac{1}{2\eta_k^2}\|\cdot - \hat\bx_{k}\|^2\right)$ using subroutine $\GDSampler(\hat\bx_{k}, \by, \cL, \eta_k)$.
    \item {\em Denoising diffusion sampler:} Sample $\hat\bx_{k+1}\sim \exp\left(  \log p^\star(x) - \frac{1}{2\eta_k^2} \|x - \hat{x}_{k + \frac12} \|^2   \right) $ using subroutine $\DiffSampler\text{-}\DDPM(\hat\bx_{k+\frac12}, \hat s, \eta_k)$ or $\DiffSampler\text{-}\DDIM(\hat\bx_{k+\frac12}, \hat s, \eta_k)$.
\end{enumerate}
\STATE \textbf{{Output}}: $\hat\bx_K$.
\end{algorithmic} 
\end{algorithm}

\paragraph{A prelude: proximal gradient method.} Informally speaking, bridging the perspective of optimization and sampling \citep{wibisono2018sampling}, posterior sampling can be viewed as a ``soft'' solution to the following optimization problem:
\begin{equation}
\label{eqn: optimization}
    \max_{\bx \in \RR^d} \; \cL(\bx; \by) + \log p^\star(\bx). 
\end{equation}
Instead of producing the point estimate that maximizes $\cL(x; y) + \log p^\star(x)$,  posterior sampling produces samples from the posterior distribution $p^\star(x)\rme^{\cL(x ; y)} = \rme^{\cL(x; y) + \log p^\star(x)}$ instead, which allows characterizing the underlying uncertainty. For better understanding, it is useful to bear in mind the special case when the image prior  $p^\star$ is supported on some low-dimensional manifold\footnote{This assumption, known as the manifold hypothesis, is commonly adopted as a flexible structural characterization of high-dimensional data. We mention it here to facilitate the understanding of our design, which will  not be imposed in our algorithm or analysis.} $\cM$. 
In this setting, we notice that $\log p^\star(\bx) = -\infty$ for $x \not\in \cM$, hence the optimization problem \eqref{eqn: optimization} is implicitly constrained in $x \in \cM$.

Recall the well-known proximal gradient method \citep{parikh2014proximal} for solving \eqref{eqn: optimization}, where one initializes a random $\hat\bx_0\in\RR^d$ and uses the following update rule
\[
    \hat\bx_{k+1} = \Prox_{\cM, \eta_k} \big( \hat\bx_k + \eta_k\nabla_{\hat\bx_k} \cL(\hat\bx_k; \by) \big), \quad k=0,1,\ldots.
\]
Here, $\eta_k>0$ is the stepsize at the $k$-th iteration, and $\Prox_{\cM,\eta}: \RR^d \to \RR^d$ is the proximal operator defined by
\begin{equation}
    \label{eqn: proximal}
    \Prox_{\cM,\eta}(x) \defeq \argmin_{x' \in \RR^d} \; -\log p^\star(x') + \frac{1}{2\eta^2} \|x' - x\|^2.
\end{equation}
Intuitively, one may view $\Prox_{\cM, \eta}(x)$ as some kind of denoising to make $x$ more consistent with its structural property \citep{venkatakrishnan2013plug}. The proximal gradient method alternatively applies two operations:
\begin{enumerate}[label=(\roman*)]
\item {\em Gradient step to enforce the measurement consistency constraint.} The gradient step tries to boost consistency $\by\approx \cA(\bx)$ via moving along the direction to increase $\cL(\cdot ; \by)$.
\item {\em Proximal mapping to enforce the data prior constraint.} The proximal step  moves the iterate towards those points that increase $\log p^\star(x)$. In particular, when $p^\star$ is supported on a low-dimensional manifold $\cM$, the proximal map forces $\bx$ to reside in $\cM$.
\end{enumerate}

\paragraph{Diffusion plug-and-play (\myalg). }
Although the proximal gradient method does not apply to the posterior sampling problem directly, we will apply the idea of alternatively enforcing these two constraints from a sampling perspective in the same spirit of \citet{vono2019split,lee2021structured,bouman2023generative}. Our algorithm,  dubbed diffusion plug-and-play (\myalg), alternates between two samplers, the denoising diffusion sampler (\DiffSampler) and the proximal consistency sampler (\GDSampler), which can be viewed as the substitutes for the proximal operator and the gradient step respectively. Given the iterate $\hat{x}_k$ and {\em annealing} parameter $\eta_k$ at the $k$-th iteration, \myalg~proceeds with the following two steps: 
\begin{enumerate}[label=(\roman*)]
\item {\em Proximal consistency sampler to enforce the measurement consistency constraint.} \myalg{} draws a sample $\hat\bx_{k+\frac12}$ from the distribution proportional to
$$ \exp\left(\cL(x \,;\, \by) - \frac{1}{2\eta_k^2}\| x - \hat\bx_{k}\|^2\right) $$
to promote the image to be consistent with the measurements. This step, which we denote as the {\em proximal consistency sampler}, can be achieved by small modifications of   standard algorithms such as Metropolis-Adjusted Langevin Algorithm (MALA) \citep{roberts1998optimal} given in Algorithm~\ref{alg:MALA}.\footnote{For practical reasons, we apply the exponential integrator to tame the discretization error of the drift term. For a derivation of this algorithm as discretization of the continuous-time Langevin dynamics, please refer to Appendix~\ref{subsec: discretizing Langevin}. } 
\item {\em Denoising diffusion sampler to enforce the data prior constraint.} \myalg{} next draws a sample  $\hat\bx_{k+1}$ from the distribution proportional to
\begin{equation}
    \label{eqn: denoising posterior factorization}
\exp\left( - \big( - \log p^\star(x) + \frac{1}{2\eta_k^2} \|x - \hat{x}_{k + \frac12} \|^2 \big)  \right)    
    \propto p^\star(x) \rme^{-\frac{1}{2\eta_k^2} \|x -  \hat{x}_{k + \frac12}\|^2}
    \propto p^\star(x^\star = x \,|\, x^\star + \eta_k w = \hat{x}_{k + \frac12})
\end{equation}
to promote the image to be consistent with the prior, where $w\sim\mathcal{N}(0,I_d)$. The last step, which follows from the Bayes' rule, makes it clear that this step can be precisely achieved by the denoising diffusion sampler (developed in Section~\ref{subsec: dds}) using solely the unconditional score function, with two options given in Algorithm~\ref{alg:DDS-DDPM} and Algorithm~\ref{alg:DDS-DDIM}.   
\end{enumerate} 
Combining both steps lead to the proposed \myalg~method described in Algorithm~\ref{alg:DPnP}.

\begin{algorithm}[t]
\caption{Proximal Consistency Sampler $\GDSampler(\bx, \by, \cL, \eta)$ (adapted from Metropolis-Adjusted Langevin Algorithm \citep{roberts1998optimal})}\label{alg:MALA} 
\begin{algorithmic} 
\STATE \textbf{{Input}}: starting point $\bx\in\RR^d$, measurements $\by\in\RR^m$, log-likelihood function of the forward model $\cL(\cdot; \by)$, proximal parameter $\eta>0$. 
\STATE \textbf{{Hyperparameter}}: Langevin stepsize $\gamma$, and the number of iterations $N$.
\STATE \textbf{{Initialization}}: $\bz_0 = \bx$. 
\STATE \textbf{{Update}}: \textbf{for} $n=0,1,\cdots,N-1$ \textbf{do} 
\begin{enumerate}[label=(\arabic*), topsep=0pt, itemsep=0em, partopsep=0pt]
    \item \textbf{{One step of discretized Langevin}}: Set $r = \rme^{-\gamma/\eta^2}$, and
    \[ \bz_{n+\frac12} = r \bz_{n} + (1 - r) \bx + \eta^2 (1 - r) \nabla_{\bz_{n}} \cL(\bz_n; \by) + \eta\sqrt{1 - r^2}\bw_n, \quad \bw_n\sim\mathcal N(0, \bI_d).
    \]
    This is equivalent to drawing $z_{n+\frac12}$ from a distribution with density $Q(\cdot; \bz_n)$, where
    \[
        Q(\bz'; \bz) = \frac{1}{(2\pi(1-r^2))^{d/2}} \exp\left(-\frac{\left\|\bz' - \left(r \bz + (1 - r) \bx + \eta^2 (1 - r) \nabla_{\bz} \cL(\bz; \by)\right)\right\|^2}{2(1-r^2)}\right).
    \]
    \item \textbf{{Metropolis adjustment}}: Compute
    \[
        q = \frac{\exp\left( \cL(\bz_{n+\frac12} ; \by) - \frac{1}{2\eta^2} \|\bz_{n+\frac12} - \bx\|^2 \right)}{\exp\left( \cL(\bz_{n} ; \by) - \frac{1}{2\eta^2} \|\bz_{n} - \bx\|^2 \right)}\cdot\frac{Q(\bz_{n}; \bz_{n+\frac12})}{Q(\bz_{n+\frac12}; \bz_{n})},
    \]
    and set
    \[
        \bz_{n+1} = \begin{cases}
            \bz_{n+\frac12}, & \text{with probability $\min(1, q)$},\\
            \bz_{n} & \text{with probability $1 - \min(1, q)$}.
        \end{cases}
    \]
\end{enumerate}
\STATE \textbf{{Output}}: $\bz_N$.
\end{algorithmic} 
\end{algorithm}

\paragraph{Discussions of \myalg.} Some comments about the proposed \myalg~method are in order. 
\begin{itemize}
\item  The proximal consistency sampler \GDSampler{} can be viewed as a ``soft'' version of the proximal point method \citep{drusvyatskiy2017proximal}. This can be seen from a first-order approximation: the maximum likelihood of the distribution $\exp\big(\cL(\cdot; \by)-\frac{1}{2\eta_k^2}\|\cdot -  \hat\bx_{k} \|^2\big)$ is attained at the point $\bx'\in\RR^d$ satisfying
\[
\nabla_{\bx'} \cL(\bx' ;\by) - \frac{1}{\eta_k^2} (\bx' - \hat\bx_{k} ) = 0,
 \quad \Longrightarrow \quad
    \bx' =  \hat\bx_{k} + \eta_k^2 \nabla_{\bx'}\cL(\bx'; \by) \approx    \hat\bx_{k} + \eta_k^2 \nabla_{ \hat\bx_{k}}\cL( \hat\bx_{k}; \by),
\]
Therefore, the proximal consistency sampler draws random samples ``concentrated'' around $\bx'$, which approximates the implicit proximal point update, akin to a gradient step at $\hat\bx_{k}$. 
\item On the other end, the denoising posterior sampler \DiffSampler{} can be regarded as a ``soft'' version of the proximal operator by comparing \eqref{eqn: denoising posterior factorization} with \eqref{eqn: proximal}. In particular, when $p^\star$ is supported on a low-dimensional manifold $\cM$, it forces $\bx$ to reside in $\cM$, like the proximal map. To see this, note that denoising posterior distribution vanishes outside $\cM$ by \eqref{eqn: denoising posterior factorization}. This is in contrast with existing algorithms \citep{romano2017little,chung2023diffusion,laumont2022bayesian,bouman2023generative,sun2023provable}, which use the conditional expectation map $\bx_{\mathsf{noisy}} \mapsto \EE_{\bx^\star\sim p^\star, \bepsilon\sim\cN(0, \bI_d)} (\bx^\star | \bx^\star + \eta \bepsilon = \bx_{\mathsf{noisy}})$ (the stepsize $\eta>0$ may vary with time) (in view of \eqref{eq:score_interpretation})  in lieu of the proximal operator for denoising. Given the presumption that the conditional expectation is a weighted average of points on $\cM$, we see that when the manifold $\cM$ is nonlinear, this map may generate points that are far apart from $\cM$, which violates the data prior constraint and hence degrades reconstruction quality.
\item We note that the proximal consistency sampler \GDSampler{} in fact admits a simple form in the special case where the forward model $\cA$ is linear, i.e. $\cA(x) = Ax$ for some matrix $A \in \RR^{m \times d}$ (assume $A$ is of full row rank), and the measurement noise $\xi \sim \cN(0, \Sigma)$ is Gaussian. In this situation, the log-likelihood function becomes
\[\cL(x; y) = -\frac12 (y-Ax)^\top \Sigma^{-1} (y-Ax) + \text{constant}.\] 
Then the target distribution of \GDSampler{} can be rewritten as a Gaussian distribution with proper mean and covariance, namely,
\[
\exp\left( \cL(\cdot; \by)-\frac{1}{2\eta_k^2}\|\cdot -  \hat\bx_{k} \|^2 \right) \propto \exp\left( -\frac12 (\cdot - \tilde x_k)^\top \tilde\Sigma_k^{-1} (\cdot - \tilde x_k) \right) = \cN(\widetilde x_k, \widetilde\Sigma_k),
\]
where
\begin{align*}
\widetilde x_k = \left(A^\top \Sigma^{-1} A + \frac{1}{\eta_k^2} I_d \right)^{-1} \left( A^\top \Sigma^{-1} y + \frac{1}{\eta_k^2} \hat\bx_k \right),
\qquad 
\widetilde \Sigma_k = \left( A^\top \Sigma^{-1} A + \frac{1}{\eta_k^2} I_d \right)^{-1}.
\end{align*}
Therefore, the proximal consistency sampler \GDSampler{} can be implemented directly by 
\[
    \hat\bx_{k+\frac12} = \GDSampler(\hat\bx_k, y, \cL, \eta_k) = \widetilde x_k + \widetilde\Sigma_k^{1/2} \bw_k, \quad \bw_k \sim \cN(0, I_d).
\]
\end{itemize}

%!TeX root = DPnP.tex

\section{Theoretical analysis}
\label{sec:analysis}

In this section, we establish both asymptotic and non-asymptotic performance guarantees of \myalg{}, whose proofs are postponed to Appendix~\ref{sec: proof asymptotic}.

\subsection{Asymptotic consistency}
 
We first collect the asymptotic correctness of our subroutines \GDSampler{} and \DiffSampler{} in the following two lemmas. The correctness of \GDSampler{} is actually well-known, see e.g., \citet[Corollary 2]{tierney1994}.
\begin{lemma}[Correctness of \GDSampler]
    \label{lem: MALA asymp}
   Under Assumption~\ref{assumption:L}, with notation in Algorithm~\ref{alg:MALA}, in the continuous-time limit:
    \[ \gamma\to0, \quad N \to \infty,\]
    the algorithm \GDSampler{} outputs samples with distribution $\propto \exp(\cL(\cdot; \by) + \frac{1}{2\eta}\|\cdot - x\|^2)$.
\end{lemma}

The next lemma guarantees the correctness of \DiffSampler{} with exact unconditional score functions.
\begin{lemma}[Correctness of \DiffSampler]
\label{lem: DDS asymp}
Assume the score function estimation $\hat s_t$ is accurate, i.e. $\hat s_t = s^\star_t$. In the continuous-time limit:
\[T\to\infty, \quad \bar\alpha_T \to 0, \quad \frac{\bar\alpha_{t-1}}{\bar\alpha_t} \to 1, \text{ unifomly in $t$},\]
both \DiffSampler-\DDIM{} and \DiffSampler-\DDPM{} output samples $x$ obeying the denoising posterior distribution $p^\star(\bx^\star =x \,|\, \bx^\star + \eta\bepsilon = \bx_{\mathsf{noisy}})$, $\bepsilon \sim \cN(0, I_d)$. 
\end{lemma}

We are now ready to state our main result, which concerns the asymptotic correctness of \myalg{}.
\begin{theorem}[Asymptotic consistency of \myalg]
\label{thm:asymp}
Under the settings of Lemma~\ref{lem: DDS asymp} and Lemma~\ref{lem: MALA asymp}, the following holds.
Let $\epsilon_1 > \epsilon_2 > \cdots$ be a decreasing sequence of positive numbers satisfying $\lim_{l\to\infty}\epsilon_l = 0$, and $0=k_0 < k_1 < k_2 < \cdots$ be an increasing sequence of integers. 
Set the annealing schedule as follows:
\begin{equation*}
    \eta_k = \epsilon_l, \quad \text{for } k_{l-1} \le k < k_l, ~l=1,2,\cdots
\end{equation*}
Let $\min_{l'=1,2,\cdots}|k_{l'} - k_{l'-1}|\to \infty$, the output $\hat x_{k_l}$ of \myalg{} converges in distribution to the posterior distribution $p^\star(\cdot|\by)$ for $l \to \infty$.
\end{theorem}

In words, Theorem~\ref{thm:asymp} establishes the asymptotic consistency of \myalg{} under fairly mild assumptions on the forward model (cf.~Assumption~\ref{assumption:L}):  as long as the sampled distributions of \DiffSampler{} and \GDSampler{} are exact, then running \myalg{} with a slowly diminishing annealing schedule of $\{\eta_k\}$ will output samples approaching the desired posterior distribution $p^\star(\cdot|\by)$ when the number of iterations $K$ goes to infinity.

\subsection{Non-asymptotic error analysis}

We now step away from the idealized setting when  the sampled distributions of \DiffSampler{} and \GDSampler{} are exact. In practice, there are many sources of errors that can influence the sampled distributions of \DiffSampler{} and  \GDSampler{}:  non-diminishing $\gamma$ and finite number of sampling steps $N$ in \GDSampler{}, and  score estimation error and finite number of discretization steps $T$ in  \DiffSampler{}. In effect, these non-idealities will make the subroutines \GDSampler{} and \DiffSampler{} \emph{inexact}. In other words, the distribution they generate will slightly deviate from the distribution they ought to sample from. In this paper, we model such deviations by the \emph{total variation} distance from the distribution generated by \GDSampler{} (resp. \DiffSampler{}) to the ideal distribution proportional to $\exp(\cL(x; y) - \frac{1}{2\eta_k^2} \| x - \hat\bx_{k} \|^2)$ (resp. $p^\star(x^\star = x | x^\star + \eta_k \epsilon = \hat\bx_{k+\frac12})$) uniformly over all iterations. Analyzing these total variations errors is out of the scope of this paper, and we point the interested readers to parallel lines of works, e.g., \citet{li2023towards,mangoubi2019nonconvex, chewi2021optimal}, among many others. In our analysis, we will assume a black-box bound for the total variation errors of \GDSampler{} and \DiffSampler{}, which can be combined with existing analyses of the respective samplers to bound the iteration complexity  of \myalg{}.
\begin{theorem}[Non-asymptotic robustness of \myalg]
\label{thm:non-asymp}
With the notation in \myalg{} (Algorithm~\ref{alg:DPnP}), set $\eta_k\equiv \eta > 0$.    Under Assumption~\ref{assumption:L}, there exists  $\lambda \coloneqq \lambda(p^\star, \cL, \eta) \in (0,1)$, such that the following holds. Define a stationary distribution $\pi_\eta$ by
\[
    \pi_\eta(x) \propto p^\star(x) q_\eta(x),
\]
where $q_\eta$ is defined by
\begin{equation}
\label{eqn: q_eta}
q_\eta(x) \defeq \rme^{\cL(\cdot ;\, y)} * p_{\eta \epsilon} (x) = \frac{1}{(2\pi)^{d/2} \eta^d} \int \rme^{\cL(x' ;\, y) - \frac{1}{2\eta^2} \|x - x'\|^2} \rmd x', \quad \epsilon \sim \cN(0, \bI_d),
\end{equation}
where $*$ denotes convolution. 
If \GDSampler{} has error at most $\epsilon_{\GDSampler}$ in total variation and \DiffSampler{} has error at most $\epsilon_{\DiffSampler}$ in total variation per iteration, then for any accuracy goal $\epsilon_{\acc}>0$, with $K \asymp \frac{\log(1/\epsilon_{\acc})}{1-\lambda}$, we have
\begin{equation}
\label{eqn: non-asymp bound}
\TV(p_{\hat\bx_K}, \pi_\eta) \lesssim \epsilon_{\acc} \sqrt{\chi^2(p_{\hat\bx_1} \,\Vert\, \pi_\eta)} + \frac{1}{1-\lambda}(\epsilon_{\DiffSampler} + \epsilon_{\GDSampler{}}) \log\left(\frac{1}{\epsilon_{\acc}}\right).
\end{equation}
\end{theorem}

Before interpreting Theorem~\ref{thm:non-asymp}, we observe that $q_0(x) = \rme^{\cL(x; y)}$, thus $\pi_0(x) \propto p^\star(x)\rme^{\cL(x; y)}$ coincides with the desired posterior distribution $p^\star(\cdot|y)$. Thus Theorem~\ref{thm:non-asymp} tells us that, assuming a constant annealing schedule $\eta_k = \eta$, the output of \myalg{} converges in total variation to the distribution $\pi_\eta$, which is a distorted version of the desired posterior distribution up to level $\eta$, with sufficiently many iterations.
 
A few remarks are in order.
\paragraph{Non-diminishing $\eta$.} It can be seen from Theorem~\ref{thm:non-asymp} that even with a nonzero $\eta$, \myalg{} already enforces the data prior strictly. On the other hand, the measurement consistency is distorted by an order of $\eta$. This is usually tolerable, since the measurements are themselves contaminated by noise, thus when $\eta$ is smaller than the noise level, the distortion would be tolerable. In practice, it is beneficial to choose an annealing schedule of $\{\eta_k\}$, which will be elaborated in Section~\ref{sec:numerical}.

\paragraph{Spectral gap and worst-case convergence rate.} The term $1-\lambda$ is known as the \emph{spectral gap} of the associated Markov chain of \myalg{}. In many situations, it can be shown that $1-\lambda \gtrsim \frac{\eta}{\operatorname{poly}(d)}$, see e.g. \citet{vono2022efficient} for the case of log-concave $p^\star$ and negative quadratic $\cL$. In such cases, the factor $\frac{1}{1-\lambda}$ in the right hand side of \eqref{eqn: non-asymp bound} can also be improved significantly \citep{altschuler2023faster}. However, under the minimal assumption in this paper and without any additional assumption on $p^\star$ and $\cL$, $1-\lambda$ can be exponentially small in the worst case, thus our result does not contradict the worst-case lower bound in \citet{gupta2024diffusion}.

\paragraph{Provable robustness.} Theorem~\ref{thm:non-asymp} indicates the performance of \myalg{} degenerates gracefully in the presence of sampling errors. To the best of our knowledge, this is the first provably consistent and robust posterior sampling method for nonlinear inverse problems using score-based diffusion priors.

%!TeX root = DPnP.tex

\section{Numerical experiments}
\label{sec:numerical}

We provide numerical evidence to corroborate the promise of \myalg{} in solving both linear and nonlinear image reconstruction tasks. We denote \myalg{} with the subroutines \DiffSampler-\DDPM{} and \DiffSampler-\DiffSampler-\DDIM{} as \myalg-\DDPM{} and \myalg-\DiffSampler-\DDIM{} respectively.

\subsection{Inverse problems}
\label{subsec: measurement ops}

We consider the following linear and nonlinear inverse problems in our experiments.

\paragraph{Phase retrieval.} We consider phase retrieval with a coded mask, which is a classical inverse problem \citep{candes2015phase}. For a $256\times 256$ image $x$ (for each color channel) in our experiments, we first generate a random mask $M \in \RR^{256 \times 256}$ (which is shared across color channels), then apply Fourier transform $\cF$ to $M \odot x$, where $\odot$ denotes the Hadamard (entrywise) product, and finally preserve only the magnitudes of the Fourier transform. Formally, the forward measurement operator is $\cA(x) = \operatorname{mag}(\cF(M \odot x))$, where $\operatorname{mag}(\cdot)$ computes the entrywise magnitude of a matrix with complex entries. The measurement noise is again set to be white Gaussian, with variance $0.2$.

\paragraph{Quantized sensing.} Quantized sensing refers to the task of reconstructing an image from its low-bit quantized version. Here, the forward measurement operator is a one-bit per channel, dithered quantization operator. More precisely, it applies entrywise the following stochastic function $Q$ with dithering level $\theta > 0$:  
\[
    Q(\mathsf{pixel}) = \begin{cases}
        1, & \text{with probability }\frac{\rme^{\mathsf{pixel} / \theta}}{1 + \rme^{\mathsf{pixel} / \theta}} \\
        -1, & \text{with probability }\frac{1}{1 + \rme^{\mathsf{pixel} / \theta}},
    \end{cases}
\]
where $\mathsf{pixel} \in [-1, 1]$ is the value of each pixel in each channel. The measurements in quantized sensing are therefore one-bit-per-channel images. The dithering level $\theta$ is set to $0.4$ in our experiments.

\paragraph{Super resolution.} The forward model for super-resolution is the bicubic downsampling operator \citep{keys1981cubic}, which is a linear operator (in fact, a block Hankel matrix). We use a downsampling ratio of $4$ in all our experiments. The measurement noise is set to be white Gaussian, with variance $0.2$.

\subsection{Experimental setups}

We compare \myalg{} with the state-of-the-art DPS algorithm
\citep{chung2023diffusion} and LGD-MC algorithm \citep{song2023loss} on the FFHQ validation dataset \citep{karras2019style} and the ImageNet validation dataset \citep{imagenet15russakovsky}. We use the same pre-trained score functions as in \cite{chung2023diffusion},\footnote{https://github.com/DPS2022/diffusion-posterior-sampling} and all images are normalized to fit into the range $[-1, 1]$.

 \begin{figure}[!th]
    \centering
    \caption{Samples of different algorithms for phase retrieval, quantized sensing, and super resolution, where  \myalg{}  generate images of higher quality and recover fine details of the image more faithfully than the state-of-the-art DPS \citep{chung2023diffusion} and LGD-MC \citep{song2023loss} algorithms.} 
    % \yc{update this figure}}
    \label{tab:pr}
    \begin{tabular}{M{0.13\textwidth} M{0.13\textwidth} M{0.13\textwidth} M{0.13\textwidth} M{0.13\textwidth} M{0.13\textwidth}}
       \toprule
        Input & DPS & LGD-MC & \myalg-\DDPM{} & \myalg-\DDIM{} & Ground truth \\
        \midrule
        \includegraphics[width=0.12\textwidth]{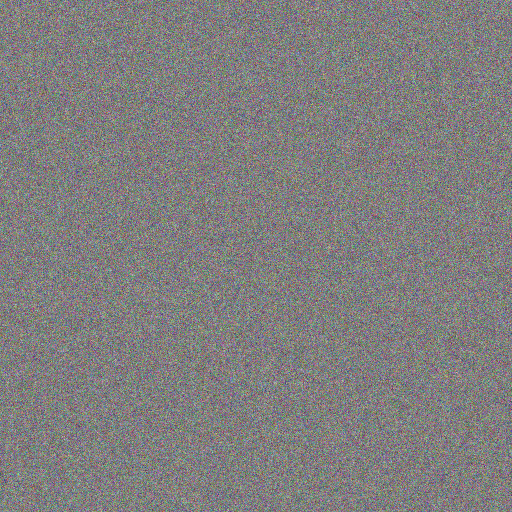} 
        & \includegraphics[width=0.12\textwidth]{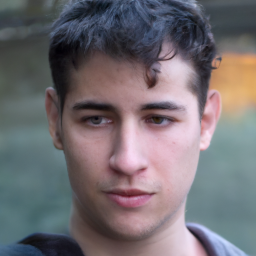} 
        & \includegraphics[width=0.12\textwidth]{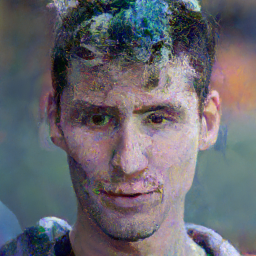} 
        & \includegraphics[width=0.12\textwidth]{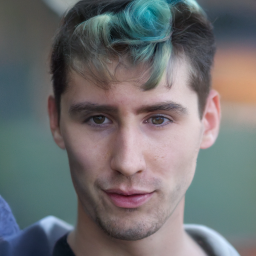} 
        & \includegraphics[width=0.12\textwidth]{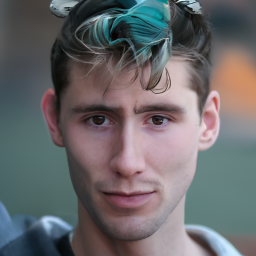} 
        & \includegraphics[width=0.12\textwidth]{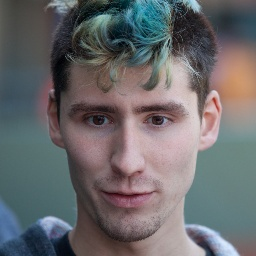} 
        \\
        \includegraphics[width=0.12\textwidth]{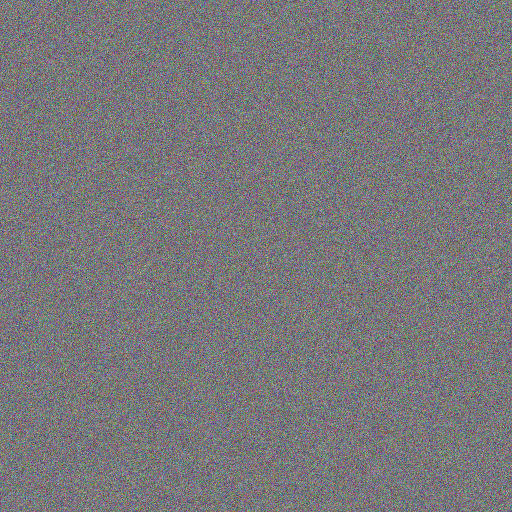} 
        & \includegraphics[width=0.12\textwidth]{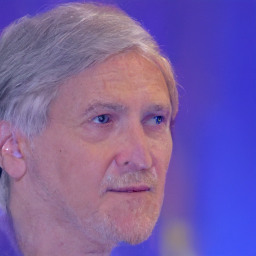} 
        & \includegraphics[width=0.12\textwidth]{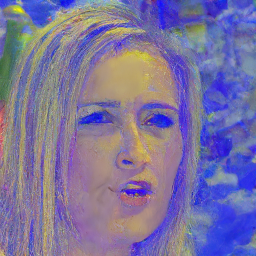} 
        & \includegraphics[width=0.12\textwidth]{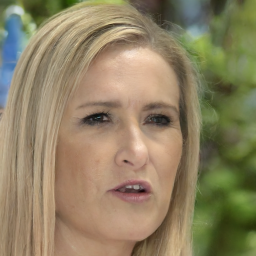} 
        & \includegraphics[width=0.12\textwidth]{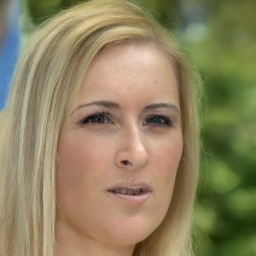} 
        & \includegraphics[width=0.12\textwidth]{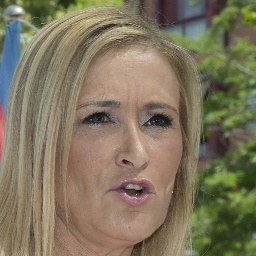} 
        \\
     %   \bottomrule
    \end{tabular} \\
    \vspace{0.05in}
    (a) phase retrieval  \\
     \vspace{0.1in}    
        \begin{tabular}{M{0.13\textwidth} M{0.13\textwidth} M{0.13\textwidth} M{0.13\textwidth} M{0.13\textwidth} M{0.13\textwidth}}
       \toprule
        Input & DPS & LGD-MC & \myalg-\DDPM{} & \myalg-\DDIM{} & Ground truth \\
        \midrule
        \includegraphics[width=0.12\textwidth]{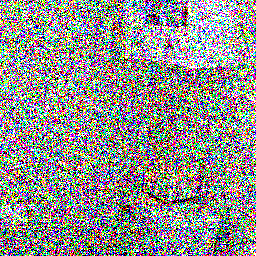} 
        & \includegraphics[width=0.12\textwidth]{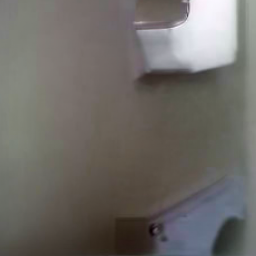} 
        & \includegraphics[width=0.12\textwidth]{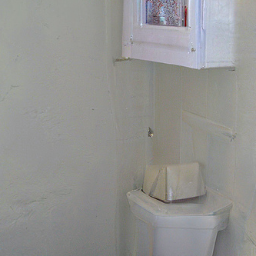} 
        & \includegraphics[width=0.12\textwidth]{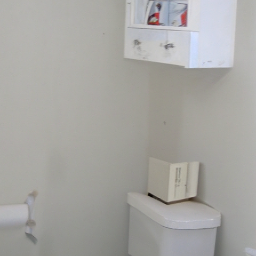} 
        & \includegraphics[width=0.12\textwidth]{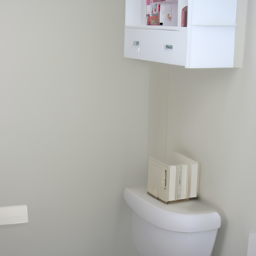} 
        & \includegraphics[width=0.12\textwidth]{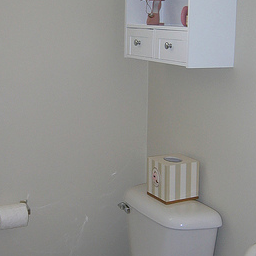} 
        \\
        \includegraphics[width=0.135\textwidth]{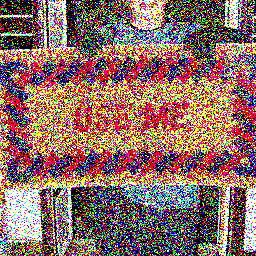} 
        & \includegraphics[width=0.135\textwidth]{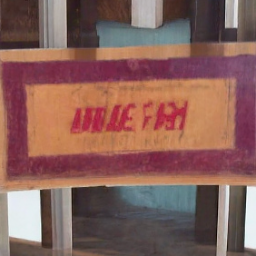} 
        & \includegraphics[width=0.135\textwidth]{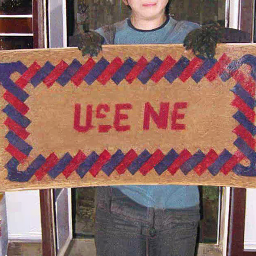} 
        & \includegraphics[width=0.135\textwidth]{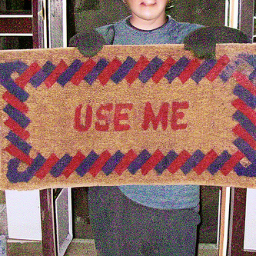} 
        & \includegraphics[width=0.135\textwidth]{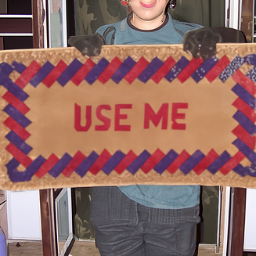} 
        & \includegraphics[width=0.135\textwidth]{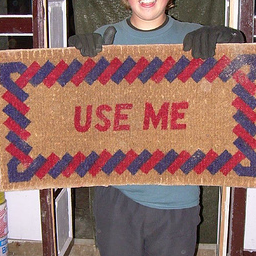} 
        \\
        \bottomrule
    \end{tabular} \\
        \vspace{0.05in}
    (b) quantized sensing  \\
     \vspace{0.1in}    
        \begin{tabular}{M{0.13\textwidth} M{0.13\textwidth} M{0.13\textwidth} M{0.13\textwidth} M{0.13\textwidth} M{0.13\textwidth}}
       \toprule
        Input & DPS & LGD-MC & \myalg-\DDPM{} & \myalg-\DDIM{} & Ground truth \\
        \midrule
        \includegraphics[width=0.12\textwidth]{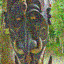} 
        & \includegraphics[width=0.12\textwidth]{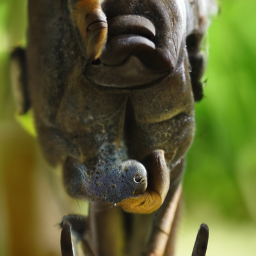} 
        & \includegraphics[width=0.12\textwidth]{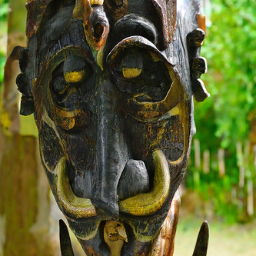} 
        & \includegraphics[width=0.12\textwidth]{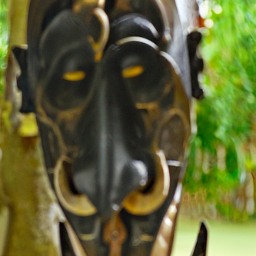} 
        & \includegraphics[width=0.12\textwidth]{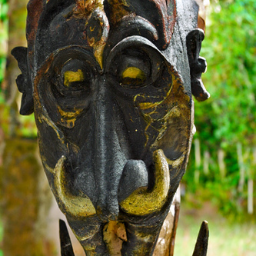} 
        & \includegraphics[width=0.12\textwidth]{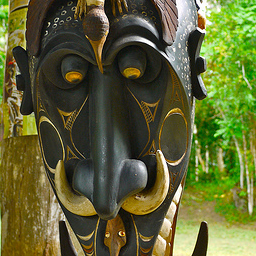} 
        \\
        \includegraphics[width=0.12\textwidth]{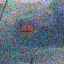} 
        & \includegraphics[width=0.12\textwidth]{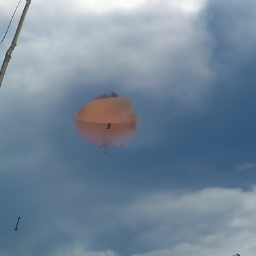} 
        & \includegraphics[width=0.12\textwidth]{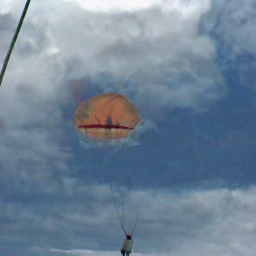} 
        & \includegraphics[width=0.12\textwidth]{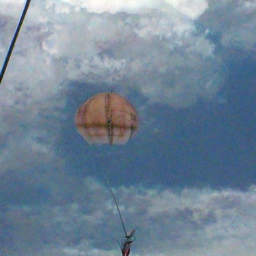} 
        & \includegraphics[width=0.12\textwidth]{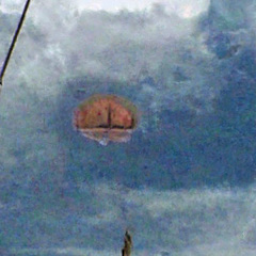} 
        & \includegraphics[width=0.12\textwidth]{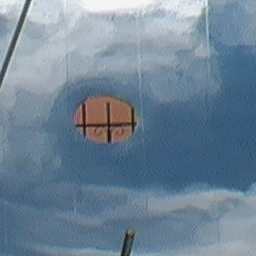} 
        \\
        \bottomrule
    \end{tabular} \\
        \vspace{0.05in} 
    (c) super resolution   
\end{figure}

\paragraph{Annealing schedule.}
For \myalg{}, we use a heuristic strategy to choose the annealing schedule $\eta_k$ in \myalg{} (Algorithm~\ref{alg:DPnP}). As seen in the theoretical analysis (Theorem~\ref{thm:non-asymp}), if we set all the $\eta_k\equiv \eta$ for some constant $\eta>0$, then \myalg{} converges to a distribution $\pi_\eta$, which can be regarded as a version of the posterior distribution $p^\star(\cdot | y)$ distorted by an order of $O(\eta)$. The smaller $\eta$ is, the more accurate the final distribution will be. On the other hand, it was also seen that in many cases, the spectral gap is $\Omega(\eta)$, hence the convergence time is $O(\frac1{\eta})$. Therefore, smaller $\eta$ would make it take longer to converge.\footnote{Strictly speaking, while the number of iterations required to converge increases as $\eta$ gets smaller, the computational complexity of \GDSampler{} and \DiffSampler{} per iteration will decrease. However, in experiments, we observed that the latter effect is not strong enough to compensate for the increase in overall complexity caused by the former.}

To strike a balance between the accuracy and the convergence rate, we adapt an gradually decreasing schedule for $\eta_k$, similar to \citet{bouman2023generative}. In the first few iterations, we set $\eta_k$ to be a large constant. After this initial phase, we decrease $\eta_k$
slowly, eventually to $\eta_N$ which is chosen to be a small constant. An example of such an annealing schedule is
\begin{align*}
    \eta_0 &= \eta_1 = \cdots = \eta_{K_0}, \quad \text{$\eta_0 > 0$ is a large constant},\\
    \eta_k &= \left(\eta_K / \eta_0\right)^{\frac{k - K_0}{K - K_0}} \eta_0, \quad K_0 < k \le K, \quad \text{$\eta_K > 0$ a small constant},
\end{align*}
where $K_0 < K$ is the length of the initial phase, which can be chosen as, e.g., $K_0 = K/5$.
For all the numerical experiments, we set $\eta_0 = 0.4$, $\eta_N=0.15$, $K_0 = 4$, $K=20$. 
The annealing schedule $\{\eta_k\}$ of \myalg{} is fixed across \emph{all} tasks, while DPS and LGD-MC are fine-tuned with reasonable effort for best performance (see Appendix~\ref{subsec:hyperparams} for the choice of hyperparameters). All experiments are run on a single Nvidia L40 GPU. %More details are in Appendix~\ref{subsec:hyperparams}.

\paragraph{Initialization.}
In Algorithm~\ref{alg:DPnP}, the initial guess $\hat x_0$ is set to be a properly scaled Gaussian random vector. Notwithstanding, from Theorem~\ref{thm:non-asymp} it can be infered that using a heuristic posterior sampler as the initializer could decrease $\chi^2(p_{\hat\bx_1} \,\Vert\, \pi_\eta)$, hence potentially improve the convergence speed of \myalg{}. By using existing algorithms like DPS or LGD-MC as initializers, \myalg{} can improve upon the results of existing algorithms towards the correct posterior distribution efficiently and provably. In our experiments, we find it helpful to initialize \myalg{} with LGD-MC, which accelerates the algorithm significantly.

\subsection{Results}

\paragraph{Visual results.} The samples generated by different algorithms are shown in Figure~\ref{tab:pr}. It can be seen from these results that, \myalg{} is capable of solving both linear and nonlinear problems, and, in comparison with state-of-the-art algorithms, performs better in recovering fine and crisper details.

\paragraph{Performance metric.}
We report the performance metric of \myalg{} in terms of LPIPS and PSNR --- which are two of the most relevant metrics for inverse problems --- on the FFHQ  and ImageNet datasets in Table~\ref{tab: ffhq} and Table~\ref{tab: imagenet}, respectively.  
Since \myalg{}-\DDIM{} has similar performance with \myalg{}-\DDPM{} but admits much faster implementation, only \myalg{}-\DDIM{} is evaluated. We refer to comparisons to other metrics such as FID and SSIM to Appendix~\ref{subsec: additional metrics}. It can be seen that \myalg{}-\DDIM{} performs strongly on both datasets, albeit taking about 1.5x more computation time.

%The LPIPS and PSNR are shown in Table~\ref{tab: ffhq} and Table~\ref{tab: imagenet}. These two metrics are arguably the more relevant ones for solving inverse problems. For comparison under other metrics such as FID, SSIM, cf. Appendix~\ref{subsec: additional metrics}. 
 
%It can be seen that, \myalg{} is capable of solving both linear and nonlinear problems, and, in comparison with prior state-of-the-art, performs better in recovering fine and crisper details. 
% The advantage of \myalg{} becomes more evident for more nonlinear forward models, e.g., phase retrieval.

 \begin{table}[ht]
  \caption{Evaluation of solving inverse problems on FFHQ $256\times 256$ validation dataset (1k samples).}
  \label{tab: ffhq}
  \centering
  \resizebox{0.95\textwidth}{!}{
  \begin{tabular}{llllllll}
    \toprule
    & \multicolumn{2}{c}{Super-resolution} & \multicolumn{2}{c}{Phase retrieval} & \multicolumn{2}{c}{Quantized sensing} &  \multicolumn{1}{c}{Time}\\
    & \multicolumn{2}{c}{(4x, linear)} & \multicolumn{2}{c}{(nonlinear)} & \multicolumn{2}{c}{(nonlinear)}  & \multicolumn{1}{c}{per sample}\\
    \cmidrule(r){2-3} \cmidrule(r){4-5} \cmidrule(r){6-7} 
    Algorithm     & LPIPS $\downarrow$ & PSNR $\uparrow$ & LPIPS $\downarrow$ & PSNR $\uparrow$ & LPIPS $\downarrow$ & PSNR $\uparrow$ & \\

    \midrule
    \myalg{}-\DDIM{} (ours) & $\mathbf{0.301}$  & $\mathbf{24.2}$ & $\mathbf{0.376}$  & $\mathbf{22.4}$ & $\mathbf{0.293}$  & $\mathbf{24.2}$    & $\sim 90$s\\
    DPS  \citep{chung2023diffusion}            & $0.331$  & $23.1$ & $0.490$  & $17.4$ & $0.367$  & $21.7$    & $\sim 60$s\\
    LGD-MC ($n=5$) \citep{song2023loss}  & $0.318$  & $23.9$ & $0.522$  & $16.4$ & $0.317$  & $23.9$ &   $\sim 60$s \\
    % FPS              & ...  & ... & n/a  & n/a & n/a  & n/a \\
    \bottomrule
  \end{tabular}
  }
\end{table}

\begin{table}[ht]
  \caption{Evaluation of solving inverse problems on ImageNet $256\times 256$ validation dataset (1k samples). }
  \label{tab: imagenet}
  \centering
  \resizebox{0.95\textwidth}{!}{
  \begin{tabular}{llllllll}
    \toprule
    & \multicolumn{2}{c}{Super-resolution} & \multicolumn{2}{c}{Phase retrieval} & \multicolumn{2}{c}{Quantized sensing} &  \multicolumn{1}{c}{Time}\\
    & \multicolumn{2}{c}{(4x, linear)} & \multicolumn{2}{c}{(nonlinear)} & \multicolumn{2}{c}{(nonlinear)}  & \multicolumn{1}{c}{per sample}\\
    \cmidrule(r){2-3} \cmidrule(r){4-5} \cmidrule(r){6-7} 
    Algorithm     & LPIPS $\downarrow$ & PSNR $\uparrow$ & LPIPS $\downarrow$ & PSNR $\uparrow$ & LPIPS $\downarrow$ & PSNR $\uparrow$ & \\

    \midrule
    \myalg{}-\DDIM{} (ours) & $\mathbf{0.416}$  & $\mathbf{21.6}$ & $\mathbf{0.562}$  & $\mathbf{13.4}$ & $\mathbf{0.363}$  & $\mathbf{23.0}$    & $\sim 240$s\\
    DPS  \citep{chung2023diffusion}            & $0.473$  & $20.2$ & $0.677$  & $\mathbf{13.4}$ & $0.542$  & $18.7$    & $\sim 150$s\\
    LGD-MC ($n=5$) \citep{song2023loss}  & $\mathbf{0.416}$  & $20.9$ & $0.592$  & $12.8$ & $0.384$  & $22.3$ &   $\sim 150$s \\
    \bottomrule
  \end{tabular}
  }
\end{table}

\section{Discussion}

\label{sec:discussion}

This paper sets forth a rigorous and versatile algorithmic framework called \myalg~for solving nonlinear inverse problems via posterior sampling, using image priors prescribed by score-based diffusion models with general forward models. \myalg~alternates between two sampling steps implemented by \DiffSampler~and \GDSampler, to promote consistency with the data prior constraint and the measurement constraint respectively. We provide both asymptotic and non-asymptotic convergence guarantees, establishing \myalg~as the first provably consistent and robust score-based diffusion posterior sampling method for general nonlinear inverse problems. Our work opens up many interesting questions, which we single out a few below.
\begin{itemize}

\item {\em Refined theory.} It is possible to further develop end-to-end finite-time convergence theory for \myalg, and obtain refined theory for structured inverse problems with additional properties of the image prior and the forward model.

\item {\em Accelerated posterior sampling.} Due to the modular design, it is straightforward to incorporated existing accelerated samplers for both \DiffSampler~\citep{lu2022dpm} and \GDSampler~\citep{ma2021there} to speed up the inference, which is of broad practical interest.

\item{\em Non-differentiable forward models.} While we assume the log-likelihood function $\mathcal{L}(\cdot; y)$ to be differentiable to apply MALA for \GDSampler, it is straightforward to adopt other samplers that only assume zero-order oracle access to $\mathcal{L}(\cdot; y)$ for non-differentiable forward models.
 
\item {\em Guided generation.} While we focus on solving inverse problems, our design might provide some insights into improving the quality of controlled or guided generation \citep{song2023loss} as well.
\end{itemize} 

We leave these directions to future work.

\section*{Acknowledgements}
 
This work is supported in part by Office of Naval Research under N00014-19-1-2404, and by National Science Foundation under DMS-2134080 and ECCS-2126634.  X. Xu is also gratefully supported by the Axel Berny Presidential Graduate Fellowship at Carnegie Mellon University.

 \bibliographystyle{apalike} 
\bibliography{reference-diffusion}

\appendix

\section{Score functions of diffusion denoising samplers}
\label{sec:der_dds}

\subsection{Proof of Lemma~\ref{lem: score of heat flow}}
\label{proof:score_hf}

\begin{proof}
The marginal distribution \eqref{eqn: solution of heat flow} of the heat flow can be written as
\begin{equation}
    \label{eqn: heat flow marginal}
    Y_\tau \eqd Y_0 + \sqrt{\tau} \epsilon, \quad Y_0\sim p^\star, ~ \epsilon\sim\cN(0, I_d).
\end{equation}
Comparing \eqref{eqn: OU marginal} and \eqref{eqn: heat flow marginal}, it is not hard to check that 
\[
Y_\tau \eqd \sqrt{1 + \tau} X_{\frac12 \log(1 + \tau)}. 
\]
Denote $\theta = \frac12 \log(1 + \tau)$ as a short-hand. We have
\[
p_{Y_\tau}(x) = p_{\sqrt{1 + \tau} X_{\theta}} (x) \propto p_{X_{\theta}} \left( \frac{1}{\sqrt{1+\tau}} x \right).
\]
Therefore it follows that
\begin{align*}
\nabla \log p_{Y_\tau}(x) = \nabla_x \log p_{X_{\theta}} \left( \frac{1}{\sqrt{1+\tau}} x \right) 
= \frac{1}{\sqrt{1+\tau}} s^\cont \left( \theta, \frac{1}{\sqrt{1+\tau}} x \right),
\end{align*}
where we used the definition $s^\cont(\theta, \cdot) = \nabla \log p_{X_\theta}(\cdot)$. Plugging the definition $\theta = \frac12 \log(1 + \tau)$ into the above equation yields the desired result.
\end{proof}

\subsection{Proof of Lemma~\ref{lem: denoising score}}
\label{proof:score_OU}

\begin{proof}
We first compute the probability density function of $\bz$. Recall that $\bz=\bw-x_{\mathsf{noisy}}$, thus applying Bayes rule yields
\begin{align*}
    p_{\bz}(\bx) = p_{\bw}(\bx + x_{\mathsf{noisy}}) = p^\star(\bx^\star = \bx + x_{\mathsf{noisy}} | \bx^\star + \xi = x_{\mathsf{noisy}}) = \frac{ p^\star(\bx + x_{\mathsf{noisy}}) p_{\xi}(-\bx) } {p_{\bx^\star + \xi}(x_{\mathsf{noisy}})}
    \propto  p^\star(\bx + x_{\mathsf{noisy}}) p_{\xi}(-\bx),
\end{align*}
where $\xi\sim\cN(0, \eta^2 \bI_d)$. It is straightforward to compute
\[
    p_{\xi}(-\bx) = \frac{1}{(2\pi)^{d/2}\eta^d} \rme^{-\frac{1}{2\eta^2} \|\bx\|^2},
\]
therefore
\begin{align}
    \label{eqn: p_z}
    p_{\bz}(\bx) \propto  p^\star(\bx + x_{\mathsf{noisy}}) \rme^{-\frac{1}{2\eta^2} \|\bx\|^2}.
\end{align}

We proceed to compute the probability density function of $\bZ_\tau$. According to \eqref{eqn:bla}, it follows that
\begin{align}
p_{\bZ_\tau}(\bx) &= p_{\rme^{-\tau} z} * p_{\sqrt{1-\rme^{-2\tau}}\bepsilon} (\bx) 
\nonumber\\
& = \int p_{\rme^{-\tau} z}(x') \, p_{\sqrt{1-\rme^{-2\tau}}\bepsilon} (x - x') \rmd x'
\nonumber\\ 
& \propto \int p_{z}(\rme^{\tau}x') \exp\left(-\frac{1}{2(1-\rme^{-2\tau})} \|x - x'\|^2\right) \rmd x'
\nonumber\\
& \propto \int p^\star( x_{\mathsf{noisy}} + \rme^{\tau}x') \exp\left(-\frac{1}{2\eta^2} \|\rme^{\tau} x'\|^2\right) \exp\left(-\frac{1}{2(1-\rme^{-2\tau})} \|x - x'\|^2\right) \rmd x', 
\nonumber\\
& \propto \int p^\star(x') \exp\left(-\frac{1}{2\eta^2} \| x' - x_{\mathsf{noisy}} \|^2\right) \exp\left(-\frac{1}{2(1-\rme^{-2\tau})} \|x - \rme^{-\tau} (x' -  x_{\mathsf{noisy}} )\|^2\right) \rmd x',
\label{eqn: p_z computation 1}
\end{align}
where $*$ denotes convolution, the penultimate line follows from \eqref{eqn: p_z} and the last line follow from the change of variable $x'\mapsto \rme^{-\tau} (x' - x_{\mathsf{noisy}} )$.
One may exercise some brute force to verify that
\begin{align*}
&\phantom{=} \exp\left(-\frac{1}{2\eta^2} \| x' - x_{\mathsf{noisy}} \|^2\right) \exp\left(-\frac{1}{2(1-\rme^{-2\tau})} \|x - \rme^{-\tau} (x' - x_{\mathsf{noisy}})\|^2\right)
\\
& = \exp \left( -\frac{\rme^{2\tau} \|x\|^2}{2(\eta^2 + \rme^{2\tau} - 1)} \right) \exp \left( -\frac{1}{2(1-\rme^{-2\tilde\tau})} \Big\| \rme^{-\tilde\tau}x_{\mathsf{noisy}} + \frac{\rme^{\tau-\tilde\tau}\eta^2 \bx}{\eta^2 + \rme^{2\tau} - 1} - \rme^{-\tilde\tau} x' \Big\|^2 \right) \\
& \propto \exp \left( -\frac{\rme^{2\tau} \|x\|^2}{2(\eta^2 + \rme^{2\tau} - 1)} \right) p_{\sqrt{1-\rme^{-2\tilde\tau}} \epsilon} \left( \rme^{-\tilde\tau}x_{\mathsf{noisy}} + \frac{\rme^{\tau-\tilde\tau}\eta^2 \bx}{\eta^2 + \rme^{2\tau} - 1} - \rme^{-\tilde\tau} x' \right),
\end{align*}
where $\tilde\tau$ is as defined in \eqref{eqn: tau prime def}. Plug this back into \eqref{eqn: p_z computation 1}, we see
\begin{align*}
p_{Z_\tau}(x) & \propto \exp \left( -\frac{\rme^{2\tau} \|x\|^2}{2(\eta^2 + \rme^{2\tau} - 1)} \right) \int p^\star(x') p_{\sqrt{1-\rme^{-2\tilde\tau}} \epsilon} \left(\rme^{-\tilde\tau}x_{\mathsf{noisy}} + \frac{\rme^{\tau-\tilde\tau}\eta^2 \bx}{\eta^2 + \rme^{2\tau} - 1} - \rme^{-\tilde\tau} x' \right) \rmd x' 
\\
& \propto \exp \left( -\frac{\rme^{2\tau} \|x\|^2}{2(\eta^2 + \rme^{2\tau} - 1)} \right) \int p^\star(\rme^{\tilde\tau} x') p_{\sqrt{1-\rme^{-2\tilde\tau}} \epsilon} \left(\rme^{-\tilde\tau}x_{\mathsf{noisy}} + \frac{\rme^{\tau-\tilde\tau}\eta^2 \bx}{\eta^2 + \rme^{2\tau} - 1} - x' \right) \rmd x' \\
& \propto \exp \left( -\frac{\rme^{2\tau} \|x\|^2}{2(\eta^2 + \rme^{2\tau} - 1)} \right) p_{\rme^{-\tau}x_0} * p_{\sqrt{1-\rme^{-2\tilde\tau}} \epsilon} \left(\rme^{-\tilde\tau}x_{\mathsf{noisy}} + \frac{\rme^{\tau-\tilde\tau}\eta^2 \bx}{\eta^2 + \rme^{2\tau} - 1}\right) \\
& \propto \exp \left( -\frac{\rme^{2\tau} \|x\|^2}{2(\eta^2 + \rme^{2\tau} - 1)} \right) p_{X_{\tilde\tau}} \left(\rme^{-\tilde\tau}x_{\mathsf{noisy}} + \frac{\rme^{\tau-\tilde\tau}\eta^2 \bx}{\eta^2 + \rme^{2\tau} - 1}\right),
\end{align*}
where the second line applies the change of variable $x'\mapsto \rme^{\tilde\tau}x'$ in the integral, the penultimate line follows from $p_{\rme^{-\tilde\tau} x_0}(x') \propto p^\star(\rme^{\tilde\tau} x')$ (since $x_0\sim p^\star$), and the last line follows from $X_{\tilde\tau} \eqd \rme^{-\tilde\tau} x_0 + \sqrt{1 - \rme^{-2\tilde\tau}} \epsilon$. 

Finally, from the above formula, we obtain
\begin{align*}
%s_{Z}(\tau, x) \defeq 
\nabla \log p_{Z_\tau}(x) 
& = \nabla_x \left( -\frac{\rme^{2\tau} \|x\|^2}{2(\eta^2 + \rme^{2\tau} - 1)} \right) + \nabla_{x} \log p_{X_{\tilde\tau}} \left(\rme^{-\tilde\tau}x_{\mathsf{noisy}} + \frac{\rme^{\tau-\tilde\tau}\eta^2 \bx}{\eta^2 + \rme^{2\tau} - 1}\right) \\
& = -\frac{\rme^{2\tau} \bx}{\eta^2 + \rme^{2\tau} - 1} + \frac{\rme^{\tau-\tilde\tau}\eta^2}{\eta^2 + \rme^{2\tau} - 1} s^\cont \left( \tilde\tau, \, \rme^{-\tilde\tau}x_{\mathsf{noisy}} + \frac{\rme^{\tau-\tilde\tau}\eta^2 \bx}{\eta^2 + \rme^{2\tau} - 1} \right),
\end{align*}
where we used the definition $s^\cont(\tilde\tau, \cdot) = \nabla \log p_{X_{\tilde\tau}}(\cdot)$.
\end{proof}

\section{Discretization with the exponential integrator}
\label{sec:discretization}

\subsection{General form of the exponential integrator}
Consider a SDE of the form:
\[
    \rmd \bM_\tau = \big( v(\tau)\bM_\tau + f(\tau, \bM_\tau)\big) \rmd \tau + \sqrt{\beta} \rmd \bB_\tau, \quad \tau\in[0, \tau_\infty], \quad \bM_0\sim p_{\bM_0},
\]
where $v: [0, \tau_\infty] \to \RR$, $f: [0,\tau_\infty]\times \RR^d \to \RR^d$ are deterministic functions, and $\beta>0$ is a constant. 
Given discretization time points $0 = \tau_0 \le \tau_1 \le \cdots \le \tau_k \le \tau_\infty$, a na\"ive way to discretize the SDE is
\[
    \bM_{\tau_{i+1}} - \bM_{\tau_i} \approx \big( v(\tau_i)\bM_{\tau_i} + f(\tau_i, \bM_{\tau_i})\big) (\tau_{i+1} - \tau_i) + \sqrt{\beta} \sqrt{\tau_{i+1} - \tau_i} \bepsilon_i, \quad i=0,1,\cdots, k-1,
\]
where $\bepsilon_i\sim\cN(0, \bI_d)$ is a standard $d$-dimensional Gaussian random vector which is independent of $\bM_{\tau_i}$. Although this approach is straightforward, it has the drawback that the linear term $v(\tau)\bM_{\tau}$ is discretized rather crude. For example, for the OU process where $v\equiv -1$, $f\equiv 0$, $\beta=2$, the SDE can be solved analytically as in \eqref{eqn: OU marginal}, while the above approach still has a discretization error. 

A more accurate discretization, known to significantly improve the quality of score-based generative models, is given by the \emph{exponential integrator} \citep{zhang2022fast}, which preserves the linear term and discretizes the SDE to
\[
    \rmd \hat\bM_\tau = \big( v(\tau)\hat\bM_\tau + f(\tau_i, \hat\bM_{\tau_i})\big) \rmd \tau + \sqrt{\beta} \rmd \bB_\tau, \quad \tau\in[\tau_i, \tau_{i+1}],\quad i=0,1,\cdots,k,
\]
with initialization $\hat\bM_0\sim p_{\bM_0}$. On each time interval $[\tau_i, \tau_{i+1}]$, this is simply a linear SDE, which can be explicitly solved by
\[
    \hat\bM_\tau \eqd \rme^{V(\tau) - V(\tau_i)} \hat\bM_{\tau_i} +  \left( \int_{\tau_i}^\tau \rme^{V(\tau)-V(\tilde\tau)} \rmd \tilde\tau \right) f(\tau_i, \hat\bM_{\tau_i}) + \sqrt{\beta} \left( \int_{\tau_i}^\tau \rme^{2(V(\tau)-V(\tilde\tau))} \rmd \tilde\tau \right)^{1/2} \bepsilon_i,
\] 
where $V$ is the antiderivative of $v$:
\[
    V(\tau) = \int_0^\tau v(\tilde\tau)\rmd \tilde\tau.
\]
Taking $\tau = \tau_{i+1}$, we obtain
\begin{align}
\hat\bM_{\tau_{i+1}} & \eqd \rme^{V(\tau_{i+1}) - V(\tau_i)} \hat\bM_{\tau_i} +\left( \int_{\tau_i}^{\tau_{i+1}} \rme^{V(\tau_{i+1})-V(\tilde\tau)} \rmd \tilde\tau \right) f(\tau_i, \hat\bM_{\tau_i}) 
\nonumber\\
& \phantom{\eqd} \quad + \sqrt{\beta} \rme^{V(\tau_{i+1})} \left( \int_{\tau_i}^{\tau_{i+1}} \rme^{2(V(\tau_{i+1})-V(\tilde\tau))} \rmd \tilde\tau \right)^{1/2} \bepsilon_i,
\label{eqn: exponential integrator solution}
\end{align}
which provides an iterative formula to compute $\hat\bM_{\tau_{i+1}}$.

%------------------------------------------------------
\subsection{Discretization of \DiffSampler-\DDPM}
\label{subsec: discretizing DDS-DDPM}
Plug the expression of $\nabla \log p_{\bY_\tau}$ in Lemma~\ref{lem: score of heat flow} into \eqref{eqn: SDE for DDS}, and use the notation $\tau^\rev = \eta^2 - \tau$, we obtain, for $\tau\in[0, \eta^2]$, that
\begin{align*}
\rmd \bY^\rev_{\tau^\rev} &= \frac{1}{\sqrt{1 + \tau^\rev}} s^\cont\!\left( \frac12 \log(1+\tau^\rev),\, \frac{\bY^\rev_{\tau^\rev}}{\sqrt{1 + \tau^\rev}} \right) \rmd \tau 
+ \rmd \tilde\bB_\tau
\\
& = -\frac{1}{\sqrt{\tau^\rev}} \epsilon^\cont\!\left( \frac12 \log(1+\tau^\rev),\, \frac{\bY^\rev_{\tau^\rev}}{\sqrt{1 + \tau^\rev}} \right) \rmd \tau 
+ \rmd \tilde\bB_\tau.
\end{align*}

\paragraph{Choosing discretization time points.}
To discretize this SDE, we first choose the discretization time points. Recalling \eqref{eqn: conversion between discrete and continuous score}, it is most reasonable to discretize at those time points $0\le \tau^\rev_0 \le \cdots \le \tau^\rev_{T'} \le \eta^2$ which satisfy
\[
\frac12 \log(1 + \tau^\rev_t) = \frac12 \log \frac{1}{\bar\alpha_t}, \quad 0\le t \le T'.
\]
This solves to
\begin{equation}
\label{eqn: DDS-DDPM discretization time}
    \tau^\rev_t = \bar\alpha_t^{-1} - 1.
\end{equation}
The requirement that $\tau^\rev_t \le \eta^2$ translates to $\bar\alpha_t \ge \frac{1}{1+\eta^2}$, which yields the following choice of $T'$: 
\begin{equation}
    \label{eqn: T prime}
    T' \coloneqq \max\left\{ t: 0\le t \le T,\,\bar\alpha_t > \frac{1}{\eta^2 + 1} \right\}.
\end{equation}
\paragraph{Applying the exponential integrator.}
Now we apply the exponential integrator to discretize the SDE on each time interval $\tau^\rev\in[\tau_{t-1}, \tau_t]$, $t=1,\cdots, T'$ as follows:
\begin{align*}
\rmd \hat\bY^\rev_{\tau^\rev} & = -\frac{1}{\sqrt{\tau^\rev}} \epsilon^\cont\!\left( \frac12 \log(1+\tau^\rev_t),\, \frac{\hat\bY^\rev_{\tau^\rev_t}}{\sqrt{1 + \tau^\rev_t}} \right) \rmd \tau 
+ \rmd \tilde\bB_\tau, 
\\
& = -\frac{1}{\sqrt{\tau^\rev}} \epsilon^\star_t\!\left(\frac{\hat\bY^\rev_{\tau^\rev_t}}{\sqrt{1 + \tau^\rev_t}} \right) \rmd \tau 
+ \rmd \tilde\bB_\tau
\\
& = -\frac{1}{\sqrt{\tau^\rev}} \epsilon^\star_t\!\left( \sqrt{\bar\alpha_t} \hat\bY^\rev_{\tau^\rev_t} \right) \rmd \tau 
+ \rmd \tilde\bB_\tau.
\end{align*}
The SDE can be integrated directly on $\tau^\rev\in[\tau_{t-1}, \tau_t]$ (see also \eqref{eqn: exponential integrator solution}, with $v\equiv0$), yielding
\begin{align*}
\hat\bY^\rev_{\tau^\rev_{t-1}} &= \hat\bY^\rev_{\tau^\rev_{t}} - 2(\sqrt{\tau^\rev_t} - \sqrt{\tau^\rev_{t-1}}) \cdot \epsilon^\star_t\!\left( \sqrt{\bar\alpha_t} \hat\bY^\rev_{\tau^\rev_t} \right) + \int_{\eta^2 - \tau_t}^{\eta^2 - \tau_{t-1}} d\tilde B_\tau \rmd \tau
\\
& \eqd \hat\bY^\rev_{\tau^\rev_{t}} - 2(\sqrt{\tau^\rev_t} - \sqrt{\tau^\rev_{t-1}}) \cdot \bepsilon^\star_t\!\left( \sqrt{\bar\alpha_t} \hat\bY^\rev_{\tau^\rev_t} \right) + \sqrt{\tau^\rev_t - \tau^\rev_{t-1}} \bw_t, 
\end{align*}
where $\bw_t \sim \cN(0, \bI_d)$ is independent of $\hat\bY^\rev_{\tau^\rev_{t}}$. Set $\hat x_t = \hat\bY^\rev_{\tau^\rev_t}$, we obtain
\begin{equation}
\label{eqn: DDS-DDPM integrator}
    \hat\bx_{t-1} \eqd \hat\bx_t - 2(\sqrt{\tau_t} - \sqrt{\tau_{t-1}}) \cdot \bepsilon^\star_{t}\!\left( \sqrt{\bar\alpha_t} \hat\bx_t \right) + \sqrt{\tau_t - \tau_{t-1}} \bw_t, \quad \bw_t \sim \cN(0, \bI_d),
\end{equation}
which is exactly the update equation in Algorithm~\ref{alg:DDS-DDPM}, except that $\bepsilon^\star_t$ is replaced by the noise estimate $\hat\epsilon_t$.

%-------------------------------------------------
\subsection{Discretization of \DiffSampler-\DDIM}
\label{subsec: discretizing DDS-DDIM}
Plug in the expression of $s_{\bZ}$ in Lemma~\ref{lem: denoising score} into the probability flow ODE~\eqref{eqn: ODE for DDS}, we obtain
\begin{align*}
\rmd \bZ^\rev_{\tau^\rev} & = \frac{\eta^2 - 1}{\eta^2 + \rme^{2\tau^\rev} - 1} \bZ^\rev_{\tau^\rev} \rmd \tau 
+ \frac{\rme^{\tau^\rev - \tilde\tau(\tau^\rev)}\eta^2}{\eta^2 + \rme^{2\tau^\rev} - 1} s^\cont\!\left( \tilde\tau(\tau^\rev), \, \rme^{-\tilde\tau(\tau^\rev)}\bx_{\mathsf{noisy}} + \frac{\rme^{\tau^\rev-\tilde\tau(\tau^\rev)}\eta^2 \bx}{\eta^2 + \rme^{2\tau^\rev} - 1} \right) \rmd \tau 
\\
& = \frac{\eta^2 - 1}{\eta^2 + \rme^{2\tau^\rev} - 1} \bZ^\rev_{\tau^\rev} \rmd \tau 
- \frac{\rme^{2\tau^\rev}}{\rme^{2\tau^\rev} - 1} \bepsilon^\cont\!\left( \tilde\tau(\tau^\rev), \, \rme^{-\tilde\tau(\tau^\rev)}\bx_{\mathsf{noisy}} + \frac{\rme^{\tau^\rev-\tilde\tau(\tau^\rev)}\eta^2 \bx}{\eta^2 + \rme^{2\tau^\rev} - 1} \right) \rmd\tau,
\end{align*}
where the second line used the definition \eqref{eqn: tau prime def}. 

\paragraph{Choosing discretization time points.}
Similar to the derivation in Appendix~\ref{subsec: discretizing DDS-DDPM}, we discretize at time points $0=\tau^\rev_0\le \tau^\rev_1\le \cdots \le \tau^\rev_{T'}\le \eta^2$, which obey
\begin{equation}
    \label{eqn: denoising ODE choice of time}
    \tilde\tau(\tau^\rev_t) = \frac12 \log\frac{1}{\bar\alpha_t}, \quad t=0,1,\ldots,T',
\end{equation}
which solves to
\begin{equation}
    \label{eqn: DDS-DDIM discretization time}
    \tau^\rev_t = \frac12 \log\frac{\eta^2 + \bar\alpha_t - 1}{(\eta^2 + 1)\bar\alpha_t - 1}.
\end{equation}
To make this well-defined, we require
\[
    \frac{\eta^2 + \bar\alpha_t - 1}{(\eta^2 + 1)\bar\alpha_t - 1} > 0,
\]
which is equivalent to
\[
    \bar\alpha_t > \frac{1}{1+\eta^2}.
\]
This leads to the same choice of $T'$ as in \eqref{eqn: T prime}. We also set
\[
    \tau_\infty = \tau^\rev_{T'}.
\]
It is convenient to introduce a notation for the corresponding discrete schedule of $\tau^\rev_t$, denoted by
\[
    \bar u_t = \rme^{-2\tau^\rev_t} = \frac{(\eta^2 + 1)\bar\alpha_t - 1}{\eta^2 + \bar\alpha_t - 1}, \quad t=0,1,\cdots,T'.
\]

\paragraph{Applying the exponential integrator.}
Now we apply the exponential integrator, which discretizes the ODE on each time interval $\tau^\rev \in [\tau_{t-1}, \tau_t]$, $t=1,\cdots,T'$, as 
\begin{align*}
\rmd \hat\bZ^\rev_{\tau^\rev} & = \frac{\eta^2 - 1}{\eta^2 + \rme^{2\tau^\rev} - 1} \hat\bZ^\rev_{\tau^\rev} \rmd \tau 
- \frac{\rme^{2\tau^\rev}}{\rme^{2\tau^\rev} - 1} \bepsilon^\cont\!\left( \tilde\tau(\tau^\rev_t), \, \rme^{-\tilde\tau(\tau^\rev_t)}\bx_{\mathsf{noisy}} + \frac{\rme^{\tau^\rev_t-\tilde\tau(\tau^\rev_t)}\eta^2 \hat\bZ^\rev_{\tau^\rev} }{\eta^2 + \rme^{2\tau^\rev_t} - 1} \right) \rmd\tau
\\
& = \frac{\eta^2 - 1}{\eta^2 + \rme^{2\tau^\rev} - 1} \hat\bZ^\rev_{\tau^\rev} \rmd \tau 
- \frac{\rme^{2\tau^\rev}}{\rme^{2\tau^\rev} - 1} \bepsilon^\star_t\!\left(\sqrt{\bar\alpha_t}\bx_{\mathsf{noisy}} + \frac{\rme^{\tau^\rev_t}\sqrt{\bar\alpha_t} \eta^2 \hat\bZ^\rev_{\tau^\rev} }{\eta^2 + \rme^{2\tau^\rev_t} - 1} \right) \rmd\tau,
\\
& = \frac{\eta^2 - 1}{\eta^2 + \rme^{2\tau^\rev} - 1} \hat\bZ^\rev_{\tau^\rev} \rmd \tau 
- \frac{\rme^{2\tau^\rev}}{\rme^{2\tau^\rev} - 1} \bepsilon^\star_t\!\left(\sqrt{\bar\alpha_t}\bx_{\mathsf{noisy}} + \frac{\sqrt{\bar u_t}\sqrt{\bar\alpha_t} \eta^2 \hat\bZ^\rev_{\tau^\rev} }{(\eta^2 - 1)\bar u_t + 1} \right) \rmd\tau,
\end{align*}
where the second line follows from \eqref{eqn: denoising ODE choice of time}, and the last line follows from dividing both the denominator and the numerator in the fraction inside $\hat\bepsilon_t$ by $\rme^{2\tau^\rev_t}$. This is a first-order linear ODE on $\tau^\rev \in [\tau_{t-1}, \tau_t]$, which can be solved explicitly (cf. \eqref{eqn: exponential integrator solution}) by 
\[
\hat\bZ^\rev_{\tau^\rev} = \frac{\sqrt{(\eta^2 - 1)\rme^{-2\tau^\rev} + 1}}{\sqrt{(\eta^2 - 1)\bar u_t + 1}} \hat\bZ^\rev_{\tau^\rev_t} + \sqrt{(\eta^2 - 1)\rme^{-2\tau^\rev} + 1} \cdot \big( h(\eta, \rme^{-2\tau^\rev}) - h(\eta, \bar u_t) \big) \cdot \bepsilon^\star_t\!\left(\sqrt{\bar\alpha_t}\bx_{\mathsf{noisy}} + \frac{\sqrt{\bar u_t}\sqrt{\bar\alpha_t} \eta^2 \hat\bZ^\rev_{\tau^\rev}}{(\eta^2 - 1)\bar u_t + 1} \right),
\]
for $\tau^\rev\in[\tau_{t-1}, \tau_t]$, where 
\[
    h(\eta, u) \defeq -\arctan\frac{\eta}{\sqrt{u^{-1} - 1}}.
\]
Plug in $\tau^\rev = \tau_{t-1}$ in the above solution, and 
set $\bz_t = \hat\bZ^\rev_{\tau^\rev_t}$, we obtain
\begin{equation}
    \label{eqn: DDS-DDIM integrator}
    \bz_{t-1} = \frac{\sqrt{(\eta^2 - 1)\bar u_{t-1} + 1}}{\sqrt{(\eta^2 - 1)\bar u_t + 1}} \bz_t + \sqrt{(\eta^2 - 1)\bar u_{t-1} + 1} \cdot \big( h(\eta, \bar u_{t-1}) - h(\eta, \bar u_t) \big) \cdot \bepsilon^\star_t\!\left(\sqrt{\bar\alpha_t}\bx_{\mathsf{noisy}} + \frac{\sqrt{\bar u_t}\sqrt{\bar\alpha_t} \eta^2 \bz_t}{(\eta^2 - 1)\bar u_t + 1} \right).
\end{equation}
The initialization, which should ideally be $z_{T'} = \hat\bZ^\rev_{\tau_\infty} \sim p_{\bZ_{\tau_\infty}}$, is approximated by $z_{T'}\sim \cN(0, \bI_d)$. 
This is exactly the update equation and the initialization in Algorithm~\ref{alg:DDS-DDIM}, except that  $\bepsilon^\star_t$ is replaced by the noise estimate $\hat\epsilon_t$.

%---------------------------------------------------------
\subsection{Discretization of \GDSampler{}}
\label{subsec: discretizing Langevin}
We first note that the Metropolis-adjustment step in \GDSampler{} (cf.~Algorithm~\ref{alg:MALA}) is standard following the classical form of MALA \citep{roberts1998optimal}. Therefore, we focus on explaining the Langevin step. Recall the continuous-time Langevin dynamics for sampling from the distribution $\exp(\cL(\cdot; \by) - \frac{1}{2\eta^2} \| \cdot - x\|^2)$:
\begin{equation}
\label{eqn:Langevin}
    \rmd \bZ_\tau = -\nabla_{\bZ_\tau} \cL(\bZ_\tau; \by) \rmd \tau + \frac{1}{\eta^2} (\bZ_\tau - \bx) \rmd \tau + \sqrt{2} \rmd \bB_\tau, \quad \tau \ge 0, \quad \bZ_0 \sim \cN(0, \bI_d).
\end{equation}
The classical form of MALA, as in \cite{roberts1998optimal}, performs one step of a straightforward discretization of \eqref{eqn:Langevin} as the Langevin step, as follows:
\[
    z_{n+\frac12} \approx z_n - \gamma \nabla_{z_n} \cL(z_n; \by) + \frac{\gamma}{\eta^2} (z_n - \bx) + \sqrt{2\gamma} \bw_n, \quad \bw_n \sim \cN(0, \bI_d).
\]
In our setting, due to the presence of the linear drift term $\frac{1}{\eta^2} (\bZ_\tau - \bx)$, which can be quite large when $\eta$ is small, we apply the exponential integrator instead. Set the discretization time points $\tau_n = n\gamma$, the exponential integrator reads as
\[
    \rmd \bZ_\tau = -\nabla_{\bZ_{n\gamma}} \cL(\bZ_{n\gamma}; \by) \rmd \tau + \frac{1}{\eta^2} (\bZ_\tau - \bx) \rmd \tau + \sqrt{2} \rmd \bB_\tau, \quad n\gamma \le \tau \le (n+1)\gamma.
\]
Solve this linear SDE on $n\gamma \le \tau \le (n+1)\gamma$ directly (see also \eqref{eqn: exponential integrator solution}) to obtain
\[
    \bZ_{(n+1)\gamma} \eqd r \bZ_{n\gamma} + (1 - r) \bx + \eta^2 (1 - r) \nabla_{\bZ_{n\gamma}} \cL(\bZ_{n\gamma}; \by) + \eta\sqrt{1 - r^2}\bw_n, \quad \bw_n\sim\mathcal N(0, \bI_d),
\]
where $r \defeq \rme^{-\gamma / \eta^2}$. This is the same as the update equation for the Langevin step in \GDSampler{} (cf.~Algorithm~\ref{alg:MALA}).

\section{Proof of main theorems}
\label{sec: proof asymptotic}

\subsection{Proof of Theorem \ref{thm:asymp}}
\begin{proof}
The proof is based on two lemmas on the one-step transition kernel of \myalg{} and the asymptotic behavior of the transition kernel, which we will present soon. 
First, we set up some notations. 
Denote
\[
    p_\eta(x) \defeq p_{x^\star\sim p^\star, \epsilon\sim\cN(0, I_d)}(x^\star + \eta\epsilon = x) = \frac{1}{(2\pi)^{d/2}\eta^d} \int p^\star(z) \rme^{-\frac{1}{2\eta^2}\|x-z\|^2} \rmd z.
\]
From the first equality, it is clear that $p_\eta \to p^\star$ when $\eta\to 0^+$. We will also use the notation $q_\eta$ defined in \eqref{eqn: q_eta}, which we recall here:
\[
    q_\eta(x) \defeq \frac{1}{(2\pi)^{d/2}\eta^d} \int \rme^{\cL(z; y) - \frac{1}{2\eta^2} \|x-z\|^2} \rmd z.
\]
In virtue of the Assumption~\ref{assumption:L}, we know that $q_\eta$ is finite for all $x \in \RR^d$.

For convenience, we introduce a notation for application of transition kernels. For a probability distribution $p(x)$ and a probability transition kernel $K(x, x')$, denote by $p \circo K$ the probability distribution given by
\[
    p \circo K(x') = \int p(x) K(x, x') \rmd x.
\]

The first lemma characterizes the one-step behavior of \myalg{} in terms of Markov transition kernels.
\begin{lemma}
\label{lem: DPnP transition kernel}
Under the settings of Lemma~\ref{lem: DDS asymp} and Lemma~\ref{lem: MALA asymp}, the one-step transition kernel of \myalg{} with $\eta_k=\eta$ is given by:
\[
    K_{\myalg, \eta}(x, x') = \left( \int \frac{q_0(z)}{p_\eta(z)} \rme^{-\frac{1}{2\eta^2}\|z-x\|^2 - \frac{1}{2\eta^2}\|z-x'\|^2} \rmd z \right) \frac{p^\star(x')}{q_\eta(x)}.
\]
In other words, if $\hat x_k$ has distribution $p_{\hat x_k}$, then the distribution of $\hat x_{k+1}$ is 
\[
    p_{\hat x_{k+1}}(x') = p_{\hat x_k} \circo K_{\myalg, \eta}(x) = \int p_{\hat x_k}(x) K_{\myalg, \eta}(x, x') \rmd x.
\]
\end{lemma}
The proof is postponed to Appendix~\ref{subsec: proof of DPnP transition kernel}.
The next lemma analyzes the ergodic properties of the Markov chain with transition kernel $K_{\myalg, \eta}$. These properties are known \citep{bouman2023generative} but scattered in different literatures, so we will provide a brief proof to be self-contained.
\begin{lemma}
\label{lem: DPnP ergodic}
The Markov transition kernel $K_{\myalg, \eta}$ has the following properties:
\begin{enumerate}[label=(\roman*)]
    \item (Stationary distribution.) Let $\pi_\eta$ be the probability distribution defined by
    \[
        \pi_\eta(x) = c_{\eta} p^\star(x) q_\eta(x),
    \]
    where $c_\eta>0$ is the normalization constant 
     such that $\int \pi_\eta(x) \rmd x = 1$. Then $K_{\myalg, \eta}$ is reversible with stationary distribution $\pi_\eta$.
    \item (Convergence.) For any initial distribution $p$, the distribution of the Markov chain with kernel $K_{\myalg, \eta}$ converges to $\pi_\eta$:
    \begin{equation}
        \label{eqn: general TV convergence}
        \TV(p \circo K_{\myalg, \eta}^{(n)},\, \pi_\eta) \to 0, \quad n\to\infty,
    \end{equation}
    where $K_{\myalg, \eta}^{(n)}$ is the $n$-step transition kernel of $K_{\myalg, \eta}$.
\end{enumerate}
\end{lemma}

The proof is postponed to Appendix~\ref{subsec: proof of DPnP ergodicity}. We now show how to prove Theorem~\ref{thm:asymp} with the above two lemmas. With the annealing schedule in Theorem~\ref{thm:asymp}, between steps $k_{l-1} \le k < k_{l}$, which consist of consecutive $(k_l - k_{l-1})$ steps, the transition kernel of one-step of $\myalg$ is $K_{\myalg, \epsilon_l}$. As $(k_l - k_{l-1})\to\infty$, Lemma~\ref{lem: DPnP ergodic} implies that
\[
    \TV(p_{\hat x_{k_l}},\, \pi_{\epsilon_l}) = \TV(p_{\hat x_{k_{l-1}}} \!\!\circo K_{\myalg, \epsilon_l}^{(k_l - k_{l-1})},\, \pi_{\epsilon_l}) \to 0.
\]
Under the assumption in Theorem~\ref{thm:asymp} that $\epsilon_l\to0$, we let $l\to\infty$ to see $\lim_{l\to \infty} \pi_{\epsilon_l} = c_0 p^\star(\cdot)\rme^{\cL(\cdot; y)} = p^\star(\cdot | y)$, thus $p_{\hat x_{k_l}}\to p^\star(\cdot | y)$, as claimed.
\end{proof}

\subsection{Proof of Lemma \ref{lem: DDS asymp}}
\begin{proof}
For \DiffSampler-\DDPM{}, we note that under the continuous-time limit in Lemma~\ref{lem: DDS asymp}, the discretization time points given by \eqref{eqn: DDS-DDPM discretization time} verify
\[
    \tau^\rev_0 = 0, 
    \quad \sup_{0 \le t \le T'-1} |\tau^\rev_t - \tau^\rev_{t+1}| \to 0, 
    \quad \tau^\rev_{T'} \to \left(\frac{1}{1+\eta^2}\right)^{-1} - 1 = \eta^2, 
    \quad T'\to\infty.
\]
Therefore, these discretization time points $0=\tau^\rev_0 \le \cdots \le \tau^\rev_{T'} \le \eta^2$ form a partition of $[0,\eta^2]$, which becomes infinitely fine in the continuous-time limit. Thus the discretized integrator \eqref{eqn: DDS-DDPM integrator} converges to the solution of the SDE \eqref{eqn: SDE for DDS}, which, as we have already argued in Appendix~\ref{sec:der_dds}, produces samples obeying the denoising posterior distribution $p^\star(\cdot | \bx_{\mathsf{noisy}})$, as claimed. 

The proof for \DiffSampler-\DDPM{} follows similarly, by observing that the discretization time points in \eqref{eqn: DDS-DDPM discretization time} form an infinitely fine partition of $[0, \infty)$ in the continuous-time limit. 
\end{proof}

%--------------------------------------------------
\subsection{Proof of Lemma \ref{lem: DPnP transition kernel}}
\label{subsec: proof of DPnP transition kernel}
\begin{proof}
The proof is based on computing the transition kernel of the two subroutines. We claim that
\begin{enumerate}[label=(\roman*)]
\item Sampling with probability density proportional to $\exp(\cL(\cdot;y) - \frac{1}{2\eta^2}\|\cdot - x\|^2)$ is equivalent to applying the following Markov transition kernel
\begin{align*}
    K_{\GDSampler, \eta}(x, x') = \frac{1}{q_\eta(x)} \rme^{\cL(x'; y) - \frac{1}{2\eta^2}\|x' - x\|^2}.
\end{align*}
\item Sampling with probability $p^\star(\bx^\star \,|\, \bx^\star + \eta\bepsilon = \bx)$, where $\bepsilon\sim\cN(0, \bI_d)$, is equivalent to applying the following Markov transition kernel:
\[
    K_{\DiffSampler, \eta}(x, x') = \frac{1}{p_\eta(x)} p^\star(x') \rme^{-\frac{1}{2\eta^2}\|x' - x\|^2}.
\]
\end{enumerate}
It is then clear that
\[
    K_{\myalg, \eta}(x, x') = \int K_{\GDSampler, \eta}(x, z) K_{\DiffSampler, \eta}(z, x')  \rmd z
    = \left( \int \frac{q_0(z)}{p_\eta(z)} \rme^{-\frac{1}{2\eta^2}\|z-x\|^2 - \frac{1}{2\eta^2}\|z-x'\|^2} \rmd z \right) \frac{p^\star(x')}{q_\eta(x)},
\]
as desired. We now prove the above two claims. For (i), note that by \eqref{eqn: denoising posterior factorization}, we know $K_{\DiffSampler, \eta}(x, \cdot) \propto p^\star(\cdot) \rme^{-\frac{1}{2\eta^2} \| \cdot - x \|^2}$. Thus it suffices to compute the normalization constant, which is
\[
    \int p^\star(x') \rme^{-\frac{1}{2\eta^2} \| x' - x \|^2} \rmd x' = p_\eta(x),
\]
by the definition of $p_\eta$. Therefore
\[
    K_{\DiffSampler, \eta}(x, x') = \frac{1}{p_\eta(x)} p^\star(x') \rme^{-\frac{1}{2\eta^2} \| x' - x \|^2},
\]
as claimed. 
The proof of (ii) follows similarly.
\end{proof}

%---------------------------------------------------
\subsection{Proof of Lemma \ref{lem: DPnP ergodic}}
\label{subsec: proof of DPnP ergodicity}
\begin{proof}
We first introduce a fundamental lemma \citep{polyanskiy2024information}, which provides a simple method to bound the total variation between two distributions.

\begin{lemma}[Data-processing inequality]
Let $p, q$ be two probability distributions, 
and $K$ be a probability transition kernel. Then 
\[
    \TV(p \circo K, q \circo K) \le \TV(p, q).
\]
\end{lemma}

We now prove the two items in Lemma~\ref{lem: DPnP ergodic} separately.

\vspace{0.5em}
\noindent{\em Proof of (i).} We first show that $\pi_\eta$ is well-defined, i.e., $\int p^\star(x) q_n\eta(x) \rmd x < \infty$. This can be seen from Assumption~\ref{assumption:L}, which implies $q_\eta(x) \lesssim \int \rme^{-\frac{1}{2\eta^2} \|x - z\|^2} \rmd z \lesssim 1$, hence
\[\int p^\star(x) q_\eta(x) \rmd x \lesssim \int p^\star(x) \rmd x = 1.\]

To show that $K_{\myalg, \eta}$ is reversible with stationary distribution $\pi_\eta$, it suffices to verify
\[
    \pi_\eta(x) K_{\myalg, \eta}(x, x') =  \pi_\eta(x') K_{\myalg, \eta}(x', x), \quad \forall x,x'\in\RR^d.
\]
However, it is easily checked that both sides are equal to
\[
    c_\eta \left( \int \frac{q_0(z)}{p_\eta(z)} \rme^{-\frac{1}{2\eta^2}\|z-x\|^2 - \frac{1}{2\eta^2}\|z-x'\|^2} \rmd z \right) p^\star(x') p^\star(x).
\]

\noindent{\em Proof of (ii).} We define an auxiliary Markov transition kernel $K_{\mathsf{aux}, \eta} = K_{\DiffSampler, \eta} \circo K_{\GDSampler, \eta}$. More explicitly,
\begin{equation}
\label{eqn: def aux kernel}
K_{\mathsf{aux}, \eta}(x, x') = \int K_{\DiffSampler, \eta}(x, z) K_{\GDSampler, \eta}(z, x') \rmd z
= \left( \int \frac{p^\star(z)}{q_\eta(z)} \rme^{-\frac{1}{2\eta^2}\|z-x\|^2 - \frac{1}{2\eta^2}\|z-x'\|^2} \rmd z \right) \frac{\rme^{\cL(x'; y)}}{p_\eta(x)}.
\end{equation}
It is easy to see that 
\begin{equation}
\label{eqn: using aux kernel}
p \circo K_{\myalg, \eta}^{(n)} = p \circo K_{\GDSampler, \eta} \circo K_{\mathsf{aux}, \eta}^{(n-1)} \circo K_{\DiffSampler, \eta}.
\end{equation}
Thus we are led to investigate the ergodic properties of $K_{\mathsf{aux}, \eta}$. Similar to the proof of item (i) above, it is not hard to show that $K_{\mathsf{aux}, \eta}$ is reversible with respect to the stationary distribution
\[
    \mu_\eta(x) \defeq  c_\eta p_\eta(x) q_0(x) = c_\eta p_\eta(x) \rme^{\cL(x; y)}.
\]
Moreover, one may check that
\begin{equation}
\label{eqn: aux stationary}
    \pi_\eta = \mu_\eta \circo K_{\DiffSampler, \eta}.
\end{equation}
It is apparent that $\mu(x') > 0$ and $K_{\mathsf{aux}, \eta}(x, x') / \mu_\eta(x') > 0$ for all $x, x'\in\RR^d$. By \citet[Corollary 1]{tierney1994}, such a Markov transition kernel obeys, for any probability distribution $q$, that
\[
    \TV( q \circo K_{\mathsf{aux}, \eta}^{(n)},~\mu_\eta) \to 0,   \quad n \to \infty.
\]
In view of \eqref{eqn: using aux kernel} and \eqref{eqn: aux stationary}, we set $q = p \circo K_{\GDSampler, \eta}$ and invoke the data-processing inequality to obtain
\begin{align*}
\TV(p \circo K_{\myalg, \eta}^{(n)}, ~ \pi_\eta)
& = \TV(q \circo K_{\mathsf{aux}, \eta}^{(n-1)} \circo K_{\DiffSampler, \eta}, ~ \mu_\eta \circo K_{\DiffSampler, \eta})
\\
& \le \TV(q \circo K_{\mathsf{aux}, \eta}^{(n-1)}, ~ \mu_\eta)
\\
& \to 0,
\end{align*}
as $n \to \infty$. This completes the proof.
\end{proof}

\subsection{Proof of Theorem \ref{thm:non-asymp}}
\begin{proof}
Denote by $\tilde K_{\GDSampler, \eta}$ and $\tilde K_{\DiffSampler, \eta}$ and the transition kernels for \GDSampler{} and for \DiffSampler{}, respectively. Note that these may deviate from the transition kernels $K_{\GDSampler, \eta}$ and $K_{\DiffSampler, \eta}$ defined for the idealized asymptotic setting in Appendix~\ref{sec: proof asymptotic}. We have
\begin{align*}
\TV(p_{\hat\bx_N}, \pi_\eta) & = \TV(p_{\hat\bx_{N-\frac12}}\!\circo \tilde K_{\DiffSampler, \eta},~\pi_\eta)
\\
& \le \TV(p_{\hat\bx_{N-\frac12}} \!\circo K_{\DiffSampler, \eta},~\pi_\eta) + \TV(p_{\hat\bx_{N-\frac12}} \!\circo K_{\DiffSampler, \eta},~ p_{\hat\bx_{N-\frac12}} \!\circo \tilde K_{\DiffSampler, \eta})
\\
& \le \TV(p_{\hat\bx_{N-\frac12}} \!\circo K_{\DiffSampler, \eta},~\pi_\eta) + \epsilon_{\DiffSampler},
\end{align*}
where the second line is triangle inequality, and the third line follows from the assumption in Theorem~\ref{thm:non-asymp} that \DiffSampler{} has error at most $\epsilon_{\DiffSampler}$ in total variation, by taking the input of \DiffSampler{} to be $\hat\bx_{N-\frac12}$. 

Similarly, from $p_{\hat\bx_{N-\frac12}} = p_{\hat\bx_{N-1}}\!\circo \tilde K_{\GDSampler, \eta}$ and the assumption that $\GDSampler{}$ has error at most $\epsilon_{\GDSampler}$ in total variation, we can show
\[
    \TV(p_{\hat\bx_{N-\frac12}} \!\circo K_{\DiffSampler, \eta},~\pi_\eta) 
    \le \TV(p_{\hat\bx_{N-1}} \!\circo K_{\GDSampler, \eta} \circo K_{\DiffSampler, \eta},~\pi_\eta) + \epsilon_{\GDSampler}
    = \TV(p_{\hat\bx_{N-1}} \!\circo K_{\myalg, \eta},~\pi_\eta) + \epsilon_{\GDSampler}.
\]
The above two inequalities together imply
\[
    \TV(p_{\hat\bx_{N}}, \pi_\eta) 
    \le \TV(p_{\hat\bx_{N-1}} \!\circo K_{\myalg, \eta},~\pi_\eta) + \epsilon_{\DiffSampler} + \epsilon_{\GDSampler}.
\]
Iterating this process, we obtain
\begin{equation}
    \label{eqn: TV bound by exact version}
    \TV(p_{\hat\bx_{N}}, \pi_\eta) 
    \le \TV(p_{\hat\bx_{1}} \circo K_{\myalg, \eta}^{(N-1)},~\pi_\eta) + (N - 1)(\epsilon_{\DiffSampler} + \epsilon_{\GDSampler}).
\end{equation}

It remains to bound $\TV(p_{\hat\bx_{1}} \circo K_{\myalg, \eta}^{(N-1)},~\pi_\eta)$. For this, we need the following two lemmas. 

\begin{lemma}[Comparing $\TV$ and $\chi^2$-divergence, \citet{polyanskiy2024information}]
For any two distributions $p, q$, we have
\[
    \TV(p, q) \le \sqrt{\chi^2(p \,\Vert\, q)}.
\]
\end{lemma}
\begin{lemma}[$\chi^2$-contractivity of $K_{\myalg,\eta}$]
\label{lem:chi2-convergence}
There exists some $\lambda \defeq \lambda(p^\star, \cL, \eta) \in (0,1)$, such that for any probability distribution $p(x)$, we have
\[
    \chi^2(p \circo K_{\myalg, \eta}^{(N)} ~\Vert ~\pi_\eta) \le \lambda^{2N} \chi^2(p \,\Vert\, \pi_\eta).
\] 
\end{lemma}
A form of Lemma~\ref{lem:chi2-convergence} is well-known for Markov chains with countable state spaces, but relatively few sources provide a complete proof for the abstract setting we consider here with continuous state space. For sake of completeness, we prove Lemma~\ref{lem:chi2-convergence} in Appendix~\ref{subsec:chi2}.

Combining the above two lemmas, we obtain
\[
    \TV(p_{\hat\bx_{1}} \circo K_{\myalg, \eta}^{(N-1)},~\pi_\eta) \le \sqrt{\chi^2(p_{\hat\bx_{1}} \circo K_{\myalg, \eta}^{(N-1)} ~\Vert~ \pi_\eta)}
    \le \lambda^{N-1} \sqrt{\chi^2(p_{\hat\bx_{1}} \,\Vert\, \pi_\eta)}.
\]
Plug this into \eqref{eqn: TV bound by exact version}, we obtain
\[
    \TV(p_{\hat\bx_{N}}, \pi_\eta) 
    \le \lambda^{N-1} \sqrt{\chi^2(p_{\hat\bx_{1}} \,\Vert\, \pi_\eta)} + (N-1) (\epsilon_{\DiffSampler} + \epsilon_{\GDSampler}).
\]
With $N\asymp \frac{\log(1/\epsilon_\acc)}{1-\lambda}$ such that $\lambda^{N - 1} \le \exp\big(-(N-1) (1-\lambda) \big) \le \epsilon_\acc$, the desired result readily follows.
\end{proof}

%-----------------------------------------------------
\subsection{Proof of Lemma \ref{lem:chi2-convergence}}
\label{subsec:chi2}
\begin{proof}
We need a few fundamental properties of reversible Markov chains, which are collected below. 

First we set up some notations. 
Define the Hilbert space $L^2(\pi)$ to be the space of square-integrable functions with respect to measure $\pi$, i.e., those functions $f : \RR^d \to \CC$ such that 
\[
    \|f\|_{L^2(\pi)} \defeq \left(\int |f(x)|^2 \pi(x) \rmd x \right)^{1/2} < \infty.
\]
The first well-known property \citep{saloff-coste1997lectures} offers a way to represent a reversible transition kernel as a self-adjoint operator (infinite-dimensional symmetric matrix).
\begin{lemma}[Self-adjoint representation of reversible Markov operator]
\label{lem: symmetrized kernel}
Assume $K(x, x')$ is a Markov transition kernel that is reversible with respect to the stationary distribution $\pi(x)$. Then the integral operator $\cK: L^2(\pi) \to L^2(\pi)$ defined by
\[
    \cK f(x) = \int K(x, x') f(x') \rmd x'
\]
is self-adjoint and compact. For any probability distribution $p(x)$ such that $\int \frac{p^2(x)}{\pi(x)} \rmd x < \infty$, we have
\[
    \int p(x) \cdot \cK f(x)  \rmd x = \int p\circo K(x') f(x') \rmd x'.
\]
Moreover, the eigenvalues of $\cK$ are the same as those of $K$.
\end{lemma}

The following theorem is a generalization of the classical Perron-Frobenius theory for finite-dimensional transition matrix to strictly positive operators. The form we present here can be found in \citet[Theorem V.6.6]{schaefer2012banach}; see also \citet[Theorem III.6.7]{bourbaki2023theories} for a more elementary treatment which can also be adapted to the form we need.
\begin{theorem}[Jentzsch]
\label{thm:eigenvalue}
Let $K(x, x')$ be a Markov transition kernel. If $K(x, x') > 0$ for any $x,x' \in \RR^d$, then $K$ has a unique stationary distribution $\pi$. Moreover, $1$ is a simple eigenvalue of $K$, with $\pi$ being the only left  eigenfunction, and the constant function $1$ being the only right eigenfunction. In addition, there exists $\lambda\in(0,1)$ such that any other eigenvalue of $K$ has modulus no larger than $\lambda$. 
\end{theorem}

We are now ready to prove Lemma~\ref{lem:chi2-convergence}. We divide the proof into the following steps.

\paragraph{Step 1: controlling the eigenvalues of $\cK_{\myalg, \eta}$.} 
Recall the auxiliary kernel $K_{\mathsf{aux}, \eta}$ defined in \eqref{eqn: def aux kernel}. It is a standard result in linear algebra or function analysis \citep{bourbaki2019theories} that $K_{\mathsf{aux}, \eta} = K_{\GDSampler, \eta} \circo K_{\DiffSampler, \eta}$ has same eigenvalues as $K_{\myalg, \eta} = K_{\DiffSampler, \eta} \circo K_{\GDSampler, \eta}$. From \eqref{eqn: def aux kernel}, it is easy to check $K_{\mathsf{aux}, \eta}(x, x') > 0$, thus Theorem~\ref{thm:eigenvalue} implies $1$ is a simple eigenvalue of $\cK_{\myalg, \eta}$. Moreover, there exists $\lambda \defeq \lambda(p^\star, \cL, \eta) \in (0,1)$, such that any other eigenvalue of $K_{\mathsf{aux}, \eta}$ has modulus no larger than $\lambda$. 

Since $K_{\myalg, \eta}$ has the same eigenvalues as $K_{\mathsf{aux}, \eta}$, and, by Lemma~\ref{lem: symmetrized kernel}, the operator $\cK_{\myalg, \eta}$ also has the same eigenvalues as these two, we conclude that $\cK_{\myalg, \eta}$ is a self-adjoint compact operator on $L^2(\pi_\eta)$, of whom $1$ is a simple eigenvalue. Moreover, any other eigenvalue of $\cK_{\myalg, \eta}$ has modulus no larger than $\lambda$.

\paragraph{Step 2: establishing the contractivity of $\cK_{\myalg, \eta}$ in $L^2(\pi_\eta)$.} 
It is easy to verify that the constant function $\mathbf 1$, which takes value $1$ for any $x \in \RR^d$, is a eigenfunction of $\cK_{\myalg, \eta}$ associated to the simple eigenvalue $1$, thus is the only (up to scaling) eigenfunction associated to that eigenvalue. It is also a unit-length eigenfunction, since $\|\mathbf 1\|_{L^2(\pi_\eta)} = (\int 1\cdot\pi_\eta(x) dx)^{1/2}=1$. Therefore, the operator $\cK_{\myalg, \eta} - \mathbf1 \mathbf1^\top$ is a self-adjoint operator whose eigenvalues have modulus no larger than $\lambda$, where $\mathbf1 \mathbf1^\top$ is the orthogonal projection onto $\mathbf 1$ in $L^2(\pi_\eta)$,  defined by
\[
    \mathbf1 \mathbf1^\top f(x) \equiv \int f(x') \pi_\eta(x') \rmd x', \quad \forall x \in \RR^d.
\]
Using the fact that $\cK_{\myalg, \eta}\mathbf1 \mathbf1^\top = \mathbf1 \mathbf1^\top\cK_{\myalg, \eta} = \mathbf1 \mathbf1^\top$, one may show $(\cK_{\myalg, \eta} - \mathbf1 \mathbf1^\top)^N = \cK_{\myalg, \eta}^{(N)} - \mathbf1 \mathbf1^\top$ by expanding the product, see e.g. \citet{saloff-coste1997lectures}. Consequently, $\cK_{\myalg, \eta}^{N} - \mathbf1 \mathbf1^\top$ is a self-adjoint operator whose eigenvalues have modulus no larger than $\lambda^N$, i.e.,
\begin{equation}
    \label{eqn: op norm bound}
    \left\| \cK_{\myalg, \eta}^N - \mathbf1 \mathbf1^\top \right\|_{L^2(\pi_\eta) \to L^2(\pi_\eta)} \le \lambda^N,
\end{equation}
where $\|\cdot\|_{L^2(\pi_\eta) \to L^2(\pi_\eta)}$ denotes the operator norm on $L^2(\pi_\eta)$. 

\paragraph{Step 3: bounding the inner product of $p \circo K_{\myalg, \eta}^{(N)} - \pi_\eta$ with any square-integrable function.} 
Note that when $\chi^2(p \,\Vert\, \pi_\eta) = \infty$, the conclusion is trivially true. For the rest part of the proof, we assume $\chi^2(p \,\Vert\, \pi_\eta) < \infty$.
Now, for any $f \in L^2(\pi_\eta)$, by applying Lemma~\ref{lem: symmetrized kernel} iteratively, we obtain
\begin{align*}
\int p \circo K_{\myalg, \eta}^{(N)}(x) f(x)  \rmd x 
& = \int p(x') \cK_{\myalg, \eta}^{N}f(x')  \rmd x'
\\
& = \int p(x') \cdot ( \cK_{\myalg, \eta}^{N} - \mathbf1 \mathbf1^\top ) f(x')  \rmd x' + \int p(x') \mathbf1 \mathbf1^\top f(x')  \rmd x'
\\
& = \int p(x') \cdot ( \cK_{\myalg, \eta}^{N} - \mathbf1 \mathbf1^\top ) f(x') \rmd x' +  \int f(x') \pi_\eta(x') \rmd x' ,
\end{align*}
where the last line follows from the definition of $\mathbf1 \mathbf1^\top$ and $\int p(x') \rmd x'=1$. Rearrange the terms to see
\begin{equation}
\label{eqn: inner product (general)}
\int \big( p \circo K_{\myalg, \eta}^{(N)}(x) - \pi_\eta(x) \big) f(x) \rmd x 
= \int p(x') \cdot ( \cK_{\myalg, \eta}^{N} - \mathbf1 \mathbf1^\top ) f(x') \rmd x'.
\end{equation}
In particular, taking $p = \pi_\eta$ yields
\begin{equation}
\label{eqn: inner product (zero)}
0 = \int \pi_\eta(x') \cdot ( \cK_{\myalg, \eta}^{N} - \mathbf1 \mathbf1^\top ) f(x') \rmd x'.
\end{equation}
Substract \eqref{eqn: inner product (zero)} from \eqref{eqn: inner product (general)}, and then take absolute value, we obtain
\begin{align}
\left| \int \big( p \circo K_{\myalg, \eta}^{(N)}(x) - \pi_\eta(x) \big) f(x) \rmd x  \right|
& = \left| \int \big( p(x') - \pi_\eta(x') \big) \cdot ( \cK_{\myalg, \eta}^{N} - \mathbf1 \mathbf1^\top ) f(x') \rmd x' \right|
\nonumber\\
& \le \left(\int \frac{\big( p(x') - \pi_\eta(x') \big)^2}{\pi_\eta(x)} \rmd x\right)^{1/2} 
\cdot \left\| ( \cK_{\myalg, \eta}^{N} - \mathbf1 \mathbf1^\top ) f(x') \right\|_{L^2(\pi_\eta)}
\nonumber\\
& \le \sqrt{\chi^2(p \,\Vert\, \pi_\eta)} \cdot \lambda^N \|f\|_{L^2(\pi_\eta)}.
\label{eqn: inner product bound}
\end{align}

\paragraph{Step 4: choosing an appropriate square-integrable function.}
Now, set 
\[
    f(x) = \frac{p \circo K_{\myalg, \eta}^{(N)}(x) - \pi_\eta(x)}{\pi_\eta(x)}.
\]
It is easily checked that
\begin{align*}
\int \big( p \circo K_{\myalg, \eta}^{(N)}(x) - \pi_\eta(x) \big) f(x) \rmd x & = \chi^2(p \circo K_{\myalg, \eta}^{(N)} ~\Vert~ \pi_\eta),
\\
\|f\|_{L^2(\pi_\eta)} & = \sqrt{\chi^2(p \circo K_{\myalg, \eta}^{(N)} ~\Vert~ \pi_\eta)}.
\end{align*}
Plug these equations into \eqref{eqn: inner product bound}, we obtain
\[
    \chi^2(p \circo K_{\myalg, \eta}^{(N)} ~\Vert~ \pi_\eta) \le \lambda^{2N} \chi^2(p \,\Vert\, \pi_\eta),
\]
as claimed.
\end{proof}

\section{Additional experiment details}
\label{sec:additional}

 %Note that the noise variance is moderately  larger than that in \cite{chung2023diffusion} to better reflect the scenario in practical inverse problems. 

\subsection{Hyperparameters of competing methods}
\label{subsec:hyperparams}
 We made our best effort to fine-tune the other algorithms within a reasonable amount of time for each task. We list the parameters, following the notation in the original papers \citep{chung2023diffusion,song2023loss}, as follows.
\begin{itemize}
    \item Phase retrieval: For DPS, the stepsize is set to $0.8$. For LGD-MC, the MC sampling variance $r_t=0.05$, the loss coefficient $\lambda=10^{-3}$, and the learning rate is set to $400.0$.
    \item Quantized sensing: For DPS, the stepsize is set to $100.0$. For LGD-MC, the MC sampling variance $r_t=0.05$, the loss coefficient $\lambda=2\times 10^{-5}$, and the learning rate is set to $500.0$.
        \item Super-resolution: For DPS, the learning rate is set to $0.6$. For LGD-MC, the MC sampling variance $r_t=0.05$, the loss coefficient $\lambda=10^{-3}$, and the learning rate is set to $60.0$.
\end{itemize}

\subsection{Additional performance metrics}
\label{subsec: additional metrics}

\paragraph{Computation time in terms of Neural Function Estimations (NFEs).} In additional to the clock time statistics, we also measure the computational cost per sample of different algorithms in terms of the number of Neural Function Estimations, i.e., the number of calls to score functions. The results are in Table~\ref{tab:NFE}. Note that the NFEs for \myalg{} depends on the initialization, the annealing schedule (and the number of timesteps for \myalg-\DDIM). We provide typical numbers of NFEs with the choice of parameters used in this paper. % given in Appendix~\ref{sec:hyperparams} and with a suitable number of timesteps.

\begin{table}[th]
    \caption{Number of NFEs for different algorithms.}
    \label{tab:NFE}
    \centering
%    \resizebox{0.7\textwidth}{!}{
    \begin{tabular}{ccccc}
      \toprule
      Algorithm & \myalg{}-\DDIM{} & \myalg{}-\DDPM & DPS & LGD-MC \\
      \midrule
      NFEs     & $\sim1500$ & $\sim3000$ & $1000$ & $1000$ \\
      \bottomrule
    \end{tabular}
  %  }
\end{table}

\paragraph{Frechet Inception Distance (FID) and Structural Similarity Index Measure (SSIM).} We also compare the FID and SSIM of different algorithms across different tasks. The results are shown in Table~\ref{tab: FFHQ additional} and Table~\ref{tab: imagenet additional}. It should be pointed out that FID is arguably not very relevant to measure the quality of solving inverse problems, as accurately solving inverse problems means that the generated distribution is close to the \emph{conditional}, i.e., \emph{posterior} distribution of the image, while FID only measure the closeness to the \emph{unconditional}, i.e., \emph{prior} distribution of the image.

\begin{table}[ht]
    \caption{FID and SSIM of solving inverse problems on FFHQ $256\times 256$ validation dataset (1k samples). }
    \label{tab: FFHQ additional}
    \centering
    \resizebox{0.9\textwidth}{!}{
    \begin{tabular}{lllllll}
        \toprule
        & \multicolumn{2}{c}{Super-resolution} & \multicolumn{2}{c}{Phase retrieval} & \multicolumn{2}{c}{Quantized sensing} \\
        & \multicolumn{2}{c}{(4x, linear)} & \multicolumn{2}{c}{(nonlinear)} & \multicolumn{2}{c}{(nonlinear)}  \\
        \cmidrule(r){2-3} \cmidrule(r){4-5} \cmidrule(r){6-7} 
        Algorithm     & FID $\downarrow$ & SSIM $\uparrow$ & FID $\downarrow$ & SSIM $\uparrow$ & FID $\downarrow$ & SSIM $\uparrow$ \\
        \midrule
        \myalg{}-\DDIM{} (ours) & $\mathbf{36.3}$  & $\mathbf{0.668}$ & $\mathbf{46.5}$  & $\mathbf{0.631}$ & $\mathbf{37.3}$  & $\mathbf{0.712}$ \\
        DPS  \citep{chung2023diffusion}            & $38.6$  & $0.636$ & $52.0$  & $0.494$ & $42.1$  & $0.601$   \\
        LGD-MC ($n=5$) \citep{song2023loss}  & $36.8$  & $0.651$ & $82.3$  & $0.414$ & $40.3$  & $0.639$  \\
        % FPS              & ...  & ... & n/a  & n/a & n/a  & n/a \\
        \bottomrule
    \end{tabular}
    }
  \end{table}

\begin{table}[H]
    \caption{FID and SSIM of solving inverse problems on ImageNet $256\times 256$ validation dataset (1k samples). }
    \label{tab: imagenet additional}
    \centering
    \resizebox{0.9\textwidth}{!}{
    \begin{tabular}{lllllll}
      \toprule
      & \multicolumn{2}{c}{Super-resolution} & \multicolumn{2}{c}{Phase retrieval} & \multicolumn{2}{c}{Quantized sensing} \\
      & \multicolumn{2}{c}{(4x, linear)} & \multicolumn{2}{c}{(nonlinear)} & \multicolumn{2}{c}{(nonlinear)}  \\
      \cmidrule(r){2-3} \cmidrule(r){4-5} \cmidrule(r){6-7} 
      Algorithm     & FID $\downarrow$ & SSIM $\uparrow$ & FID $\downarrow$ & SSIM $\uparrow$ & FID $\downarrow$ & SSIM $\uparrow$ \\
      \midrule
      \myalg{}-\DDIM{} (ours) & $47.5$  & $\mathbf{0.510}$ & $\mathbf{73.5}$  & $0.289$ & $\mathbf{43.2}$  & $\mathbf{0.623}$ \\
      DPS  \citep{chung2023diffusion}            & $61.4$  & $0.496$ & $92.7$  & $\mathbf{0.318}$ & $82.4$  & $0.459$   \\
      LGD-MC ($n=5$) \citep{song2023loss}  & $\mathbf{46.2}$  & $0.503$ & $89.8$  & $0.234$ & $46.8$  & $0.563$  \\
      % FPS              & ...  & ... & n/a  & n/a & n/a  & n/a \\
      \bottomrule
    \end{tabular}
    }
  \end{table}

%\newpage

%\input{auxiliary}

\end{document}